\documentclass[acmsmall,nonacm]{acmart}
\usepackage{textcomp}
\usepackage{xcolor}
\usepackage{subcaption}
\usepackage{array}
\usepackage{makecell}
\usepackage{multirow}
\usepackage{balance}

\AtBeginDocument{%
  }

\setcopyright{acmlicensed}
\copyrightyear{2018}
\acmYear{2018}
\acmDOI{XXXXXXX.XXXXXXX}

\acmJournal{JACM}
\acmVolume{37}
\acmNumber{4}
\acmArticle{111}
\acmMonth{8}

\begin{document}

\title{Demystifying Starlink Network Performance under Vehicular Mobility with Dynamic Beam Switching}

\author{Jinwei Zhao}
\authornotemark[1]
\authornote{Jinwei Zhao and Jack Baude contributed equally to this work.}
\email{clarkzjw@uvic.ca}
\affiliation{%
  \institution{University of Victoria}
  \city{Victoria}
  \state{BC}
  \country{Canada}
}

\author{Jack Baude}
\authornotemark[2]
\authornotemark[1]
\email{baude022@umn.edu}
\affiliation{%
  \institution{University of Minnesota -- Twin Cities}
  \city{Minneapolis}
  \state{MN}
  \country{USA}
}

\author{Ali Ahangarpour}
\authornotemark[1]
\email{aliahangarpour@uvic.ca}
\affiliation{%
  \institution{University of Victoria}
  \city{Victoria}
  \state{BC}
  \country{Canada}
}

\author{Vaibhava Krishna Devulapalli}
\authornotemark[2]
\email{divak014@umn.edu}
\affiliation{%
  \institution{University of Minnesota -- Twin Cities}
  \city{Minneapolis}
  \state{MN}
  \country{USA}
}

\author{Sree Ganesh Lalitaditya Divakarla}
\authornotemark[2]
\email{devul009@umn.edu}
\affiliation{%
  \institution{University of Minnesota -- Twin Cities}
  \city{Minneapolis}
  \state{MN}
  \country{USA}
}

\author{Zhi-Li Zhang}
\authornotemark[2]
\email{zhzhang@cs.umn.edu}
\affiliation{%
  \institution{University of Minnesota -- Twin Cities}
  \city{Minneapolis}
  \state{MN}
  \country{USA}
}

\author{Jianping Pan}
\authornotemark[1]
\email{pan@uvic.ca}
\affiliation{%
  \institution{University of Victoria}
  \city{Victoria}
  \state{BC}
  \country{Canada}
}

\renewcommand{\shortauthors}{Zhao et al.}

\begin{abstract}
\normalsize

In the last few years, considerable research efforts have focused on measuring and improving Starlink network performance, especially for user terminals (UTs) in stationary scenarios.
However, the performance of Starlink networks in mobility settings, particularly with frequent changes in the UT's orientation, and the impact of environmental factors, such as transient obstructions, has not been thoroughly studied, leaving gaps in understanding the causes of performance degradation.
Recently, researchers have started identifying the communicating satellites to evaluate satellite selection strategies and the impact on network performance.
However, existing Starlink satellite identification methods only work in stationary, obstruction-free scenarios, as they do not account for UT mobility, obstructions or detect dynamic beam switching events.
In this paper, we reveal that the UT can perform multiple dynamic beam switching attempts to connect to different satellites when the UT-satellite link is degraded.
This degradation can occur either due to the loss of line-of-sight (LoS) from changes in the FOV or obstructions, or due to poor signal quality, extending UT-satellite handovers beyond the well-known 15-second regular handover interval.
We propose a mobility-aware Starlink satellite identification method that detects dynamic beam switching events, and plausibly explain network performance using UT's diagnostic data and connected satellite information.
Our findings demystifies the mobile Starlink network performance degradations, which is crucial to enhance the end-to-end performance of transport layer protocols and in diverse application scenarios.

\end{abstract}



\makeatletter
\let\@authorsaddresses\@empty
\makeatother

\maketitle

\section{Introduction}\label{sec:introduction}

The advent of low-earth-orbit (LEO) satellite networks such as SpaceX's Starlink, which has the largest LEO satellite constellation with over 9,000 operational satellites in orbit as of December 2025~\cite{celestrakNORADGPElement}, has brought high-speed, low-latency Internet to remote and underserved areas.
The majority of Starlink UTs are installed at fixed locations to serve residential or non-mobile business users.
Beyond fixed-site installations, Starlink UTs are increasingly deployed in various mobility scenarios, including on cars, recreational vehicles (RVs), trucks, trains, marine vessels, and airplanes.
When mounted on vehicles, the UTs are typically installed at fixed tilt angles.
However, their orientation frequently changes due to vehicular movements.
Furthermore, transient obstructions from different environments such as bridges, highway signs, and dense foliage coverage in rural areas, frequently block the line-of-sight (LoS) between UTs and communicating satellites, degrading the received signal-to-noise ratio (SNR) and causing network performance deterioration.

Existing studies~\cite{panMeasuringSatelliteLinks2024b,tanveerMakingSenseConstellations2023c,izhikevichDemocratizingLEOSatellite2024b,ahangarpourTrajectorybasedServingSatellite2024b} have revealed that Starlink performs regular UT-satellite handovers every 15 seconds, specifically at the 12th, 27th, 42nd, and 57th (12-27-42-57) seconds of every minute, synchronized globally.
Unlike other LEO operators such as OneWeb, Starlink does not provide users with the communicating satellite identifiers (IDs)~\cite{zhaoMeasuringOneWebSatellite2025}.
Researchers have started utilizing obstruction maps to identify communicating satellites for Starlink UTs in stationary scenarios~\cite{izhikevichDemocratizingLEOSatellite2024b,tanveerMakingSenseConstellations2023c,ahangarpourTrajectorybasedServingSatellite2024b,liuVivisectingStarlinkThroughput2025}.
Recently, Starlink issued a public statement confirming the existence of dynamic beam switching events when UTs are obstructed or under mobility scenarios~\cite{starlinkStarlinkBeamSwitching}.
This capability is enabled by the increased satellite density in most regions, with abundant visible satellites within the FOV fundamentally changing the handover problem in satellite networks.
Rather than being limited to rigid, schedule-driven transitions~\cite{chenLowLatencyScheduledriven2024a}, the network can now support dynamic, performance-driven handovers across satellites and ground stations, bringing satellite systems closer to the rich handover design space long studied in terrestrial networks.
In contrast, sparser constellations lack sufficient instantaneous alternatives, leaving little room for reactive or optimization-driven handover decisions.

In this paper, we propose a mobility-aware method to identify communicating satellites for Starlink UTs in motion, accounting for unstable alignment caused by vehicular movements.
With the detection of dynamic beam switching events, we effectively identify communicating satellites under different obstruction and mobility conditions.
Our observations on stationary obstructed UTs indicate that, when the SNR degradation occurs, UTs can make multiple dynamic beam switching attempts to recover the link and restore the network connectivity.
Mobile UTs are more likely to encounter transient obstructions than stationary UTs, leading to more frequent dynamic beam switching events and network interruptions.

In this paper, we aim to answer the following questions:

\begin{itemize}
    \item \textbf{Q1}: How do dynamic beam switching events impact network performance, including latency and throughput, and how does UT recover from network disruptions?
    \item \textbf{Q2}: What is Starlink UT's satellite selection strategy with deteriorating perceived SNR?
    \item \textbf{Q3}: How do different environmental scenarios such as varying obstruction conditions impact the network performance of Starlink UTs?
\end{itemize}

The main contributions and findings of the paper are summarized as follows:

\begin{itemize}
    \item We propose a practical mobility-aware method to identify communicating satellites for Starlink UTs, which accounts for vehicular movements and detects dynamic beam switching events. This method supports both stationary and mobile scenarios and operates effectively across various obstruction conditions (\S\ref{sec:mobile-identification}).
    \item To the best of our knowledge, we are the first to demonstrate that multiple dynamic beam switching attempts may be triggered within a 15-second timeslot to circumvent obstructions or poor SNR. We provided concrete explanations for network performance degradation with connected satellite information and UT's diagnostic data in mobility scenarios. (\S\ref{sec:mobile-identification}, \S\ref{sec:analysis})
    \item Our findings indicate that when a stationary UT is obstructed, despite having the dynamic beam switching capability, multiple failed beam switching attempts may happen, causing prolonged network interruptions. The proactive beam switching capability on stationary UTs can drastically reduce the obstruction time ratio over extended time periods. (\S\ref{sec:analysis}).
    \item We have found that mobile Starlink UTs encounter transient obstructions and consequently perform reactive beam switching events more frequently. This can result in a doubling of \texttt{SKY\_SEARCH} events, accounting for up to 45.18\% of the total outage time, compared to a stationary UT that is heavily obstructed by trees with a 17.9\% FOV obstruction ratio (\S\ref{sec:analysis}).
\end{itemize}

\section{Background}\label{sec:background}

\begin{figure}
    \centering
    \begin{subfigure}[h]{0.24\linewidth}
        \includegraphics[width=\linewidth]{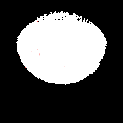}
        \caption{Stationary UT, no obstruction}
        \label{subfig:stationary-obstruction-free}
    \end{subfigure}
    \begin{subfigure}[h]{0.24\linewidth}
        \includegraphics[width=\linewidth]{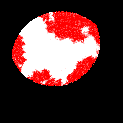}
        \caption{Stationary UT, heavily obstructed}
        \label{subfig:stationary-obstructed}
    \end{subfigure}
    \begin{subfigure}[h]{0.24\linewidth}
        \includegraphics[width=\linewidth]{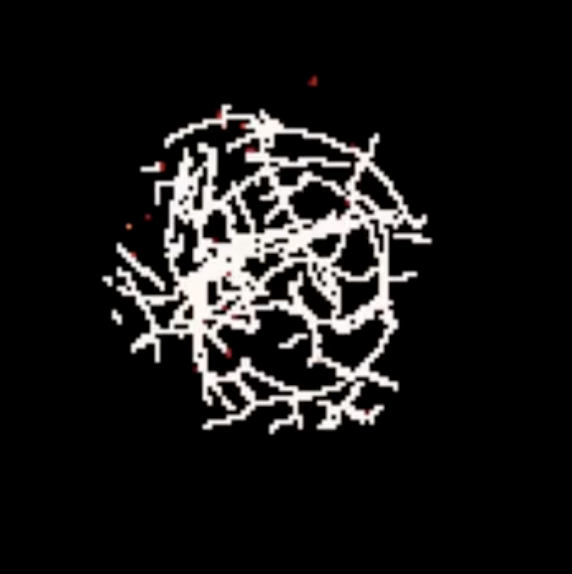}
        \caption{Mobile UT, intermediate map}
        \label{subfig:mobile-intermediate}
    \end{subfigure}
    \begin{subfigure}[h]{0.24\linewidth}
        \includegraphics[width=\linewidth]{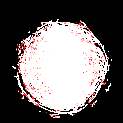}
        \caption{Mobile UT, full cumulative map}
        \label{subfig:mobile-cumulatived}
    \end{subfigure}
    \caption{Starlink UT obstruction maps obtained from the gRPC interface in different scenarios. \textbf{Black:} Unexplored region. \colorbox{black}{\textcolor{white}{White:}} Obstruction-free FOV. \textcolor{red}{\textbf{Red}:} Obstructed area.}
    \label{fig:stationary-dish-obstruction}
\end{figure}

\subsection{Starlink UT Obstruction Maps}\label{subsec:background-obstruction-maps}

Starlink UT's built-in \textbf{g}RPC \textbf{R}emote \textbf{P}rocedure \textbf{C}alls (gRPC) interface provides various diagnostic information to users, such as the UT's orientation, GPS location, and network performance statistics.
It does not reveal raw antenna metrics such as the received signal strength indicator (RSSI), SNR, or communicating satellite IDs.
Internally, the UT uses the RSSI and SNR to assess the signal quality of the communicating satellites at specific beam steering angles~\cite{mccormickDeterminationElectronicBeam2024,robinsonApparatusesMethodsFacilitating2024a}.
If the SNR drops below a certain threshold, the UT identifies obstructions in the designated direction.
The UT projects the azimuth and elevation of communicating satellites on a two-dimensional (2D) obstruction map, which illustrates the accumulated trajectories of connected satellites over time.
Give the dense constellation of Starlink satellites and the compact 123x123 pixel resolution of the obstruction map, changes in the communicating satellite trajectories are visible in pixel coordinates, but with limited temporal resolution~\cite{tanveerMakingSenseConstellations2023c,ahangarpourTrajectorybasedServingSatellite2024b,izhikevichDemocratizingLEOSatellite2024b,liuVivisectingStarlinkThroughput2025}.

Previously, Starlink UTs clear and reconstruct obstruction maps upon boot.
Early researchers have to frequently reboot the UT to clear the obstruction map to avoid overlapping trajectories~\cite{tanveerMakingSenseConstellations2023c,izhikevichDemocratizingLEOSatellite2024b}.
Starlink now adds the capability to allow users to manually reset the obstruction map and initiate the reconstruction process via gRPC commands or through the Starlink mobile application, for instance, when the physical orientation of the UT has significantly changed or the UT is relocated to a new location.
It typically takes a few hours for the obstruction map to fully converge, at which point the entire FOV is covered by the trajectories of communicating satellites, as shown in Fig.~\ref{subfig:stationary-obstruction-free} and Fig.~\ref{subfig:stationary-obstructed}.
The UT in Fig.~\ref{subfig:stationary-obstruction-free} is mounted on a building rooftop, with a clear LoS of the sky, resulting in an obstruction-free map.
In Fig.~\ref{subfig:stationary-obstructed}, the UT is situated in a residential neighborhood surrounded by trees, resulting in an FOV obstruction rate of 17.9\%, as reported by the UT's gRPC interface.

When the UT is mounted on a vehicle, unlike cameras, the obstruction map available through the Starlink mobile application or directly obtained from the gRPC interface does not provide a real-time reflection of obstructions across the entire FOV.
It is updated only with the trajectories of communicating satellite between successive frames.
For instance, Fig.~\ref{subfig:mobile-intermediate} shows an intermediate obstruction map captured during driving, not long after the obstruction map was initially cleared.
Due to the constraints of a 2D representation without temporal details, over time, the cumulative obstruction map gradually fills up as shown in Fig.~\ref{subfig:mobile-cumulatived}.
It is unable to convey meaningful information about transient obstructions at specific times along different sections of the route.
The accumulated satellite trajectories in obstruction maps are originally intended to help users identify structural obstructions in the LoS for fixed-site installations.
In this paper, we integrate the obstruction map with UT mobility and utilize temporal information to develop a mobility-aware satellite identification method for mobile Starlink UTs.

\subsection{Reference Frames in Obstruction Maps}\label{subsec:background-reference-frames}

\begin{figure}
    \centering
    \begin{subfigure}[c]{0.178\linewidth}
        \centering
        \includegraphics[width=\linewidth]{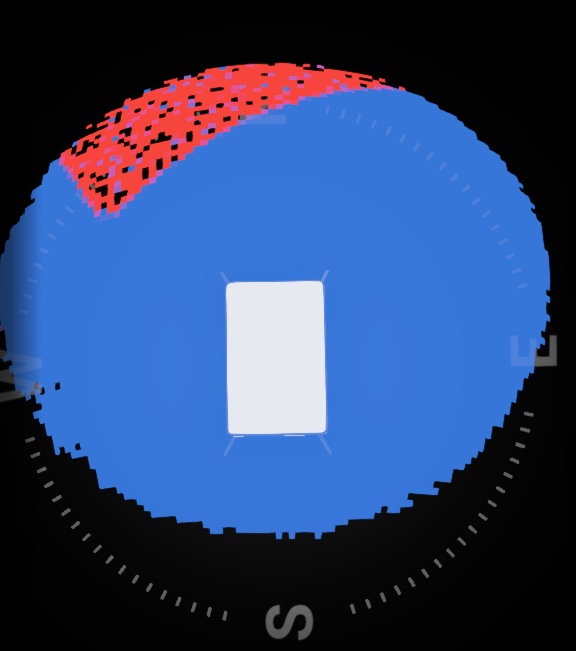}%
        \caption{Mobile app}%
        \label{fig:mobile_app}
    \end{subfigure}
    \begin{subfigure}[c]{0.2\linewidth}
        \centering
        \includegraphics[width=\linewidth]{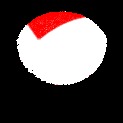}%
        \caption{\texttt{FRAME\_EARTH}}%
        \label{fig:frame_earth}
    \end{subfigure}
    \begin{subfigure}[c]{0.2\linewidth}
        \centering
        \includegraphics[width=\linewidth]{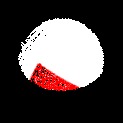}%
        \caption{\texttt{FRAME\_UT}}%
        \label{fig:frame_ut}
    \end{subfigure}

    \begin{subfigure}[c]{0.295\linewidth}
        \centering
        \includegraphics[width=\linewidth]{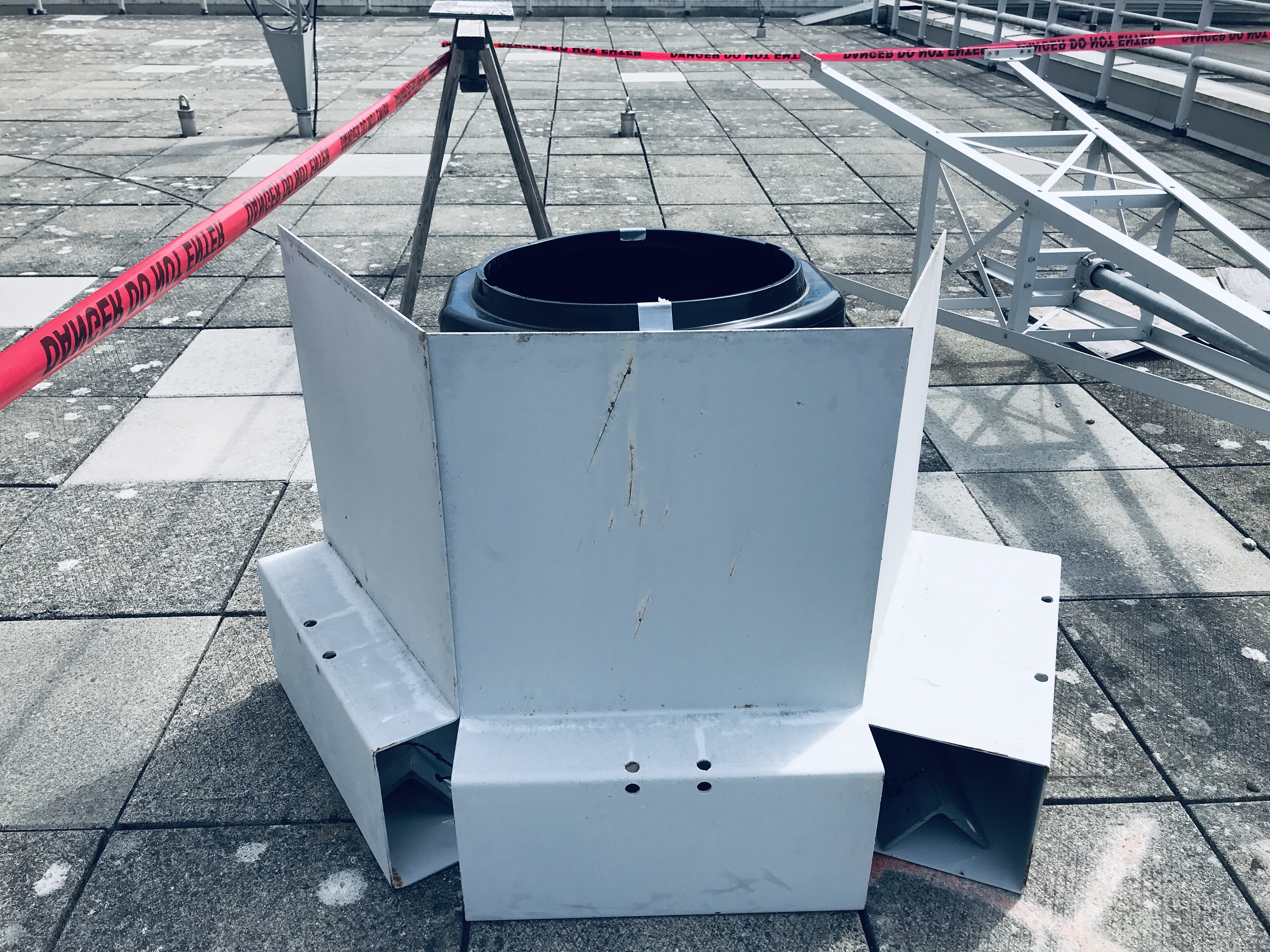}%
        \caption{Purposely obstructed UT}%
        \label{fig:frame_ut_purpose_obstructed_photo}
    \end{subfigure}
    \begin{subfigure}[c]{0.295\linewidth}
        \centering
        \includegraphics[width=\linewidth]{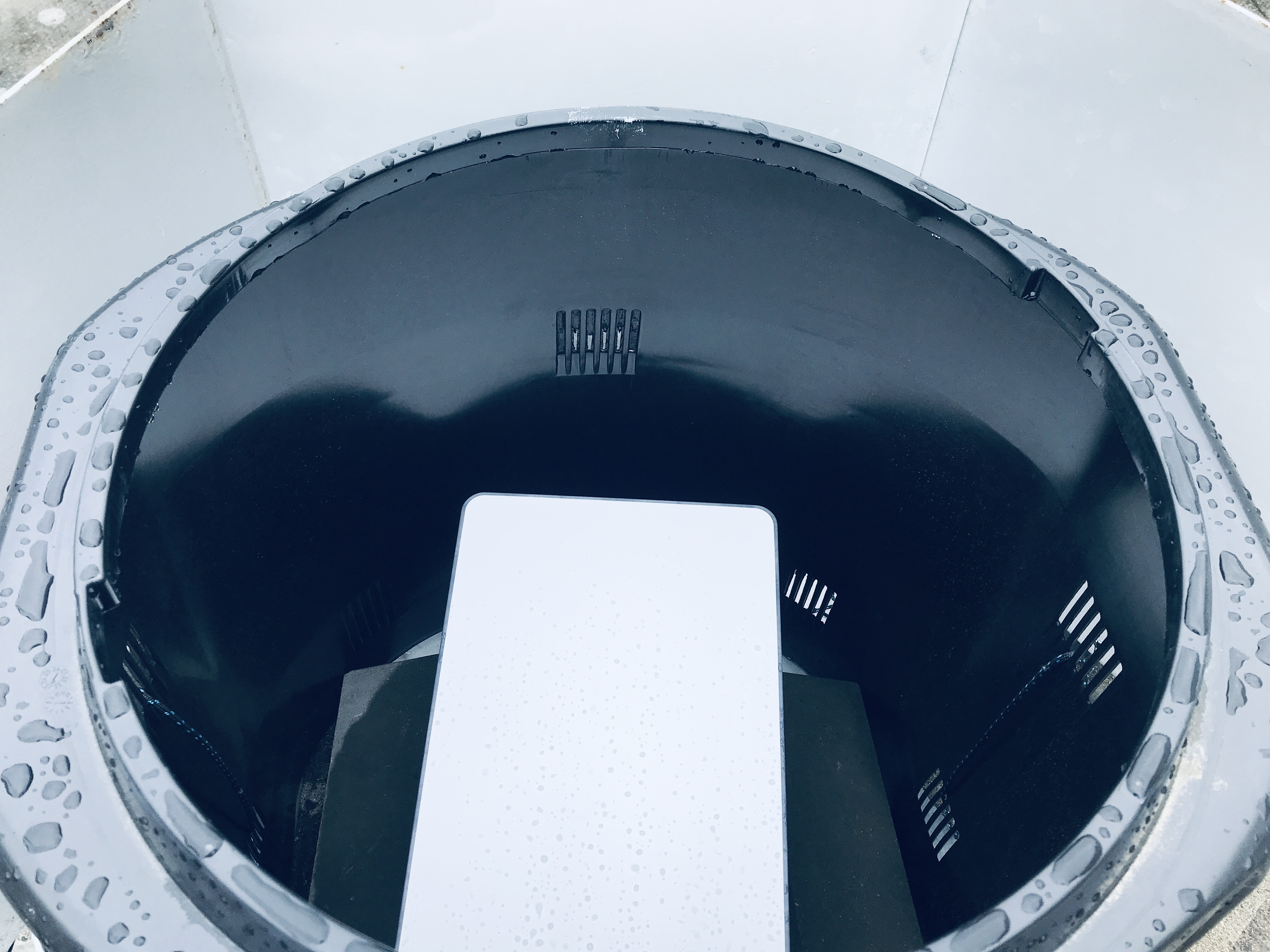}%
        \caption{The UT is set a tilt of 7.7$^\circ$}%
        \label{fig:frame_ut_purpose_obstructed_photo_2}
    \end{subfigure}
    \begin{subfigure}[c]{0.22\linewidth}
        \centering
        \includegraphics[width=\linewidth]{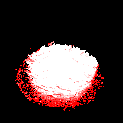}%
        \caption{Obstruction map}%
        \label{fig:frame_ut_purpose_obstructed}
    \end{subfigure}
    \caption{Obstructed Starlink UT obstruction maps presented in different reference frames, as viewed within the mobile application and retrieved from the gRPC interface. (a)-(c): UT A; (d)-(f): UT B. Both UTs share the same \texttt{rev3\_proto2} hardware model.}%
    \label{fig:obstruction_maps}
\end{figure}

Starlink UT utilizes phased array antennas to steer beams and maintain connectivity with communicating satellites~\cite{mccormickDeterminationElectronicBeam2024}.
To accurately track communicating satellites through beam steering, the UT requires a reference frame to determine its orientation.
Since firmware update in September 2024, the 2D obstruction maps obtained through the gRPC interface are represented in two different reference frames, depending on the user's subscription plan for the UT, as illustrated in Fig.~\ref{fig:mobile_app} to Fig.~\ref{fig:frame_ut}.
Such information is indicated by the \texttt{mapReferenceFrame} field in the gRPC response of the \texttt{dish\_get\_obstruction\_map} method call.

For UTs subscribed to the residential and other fixed-site plans, the reference frame (\texttt{FRAME\_EARTH}) is based on the Earth-Centered, Earth-Fixed (ECEF) reference system.
Inactive UTs, even without active subscriptions, can still access the satellite link and reach certain Internet resources, including Starlink's website and the billing system~\cite{panMeasuringSatelliteLinks2024b}.
Both inactive UTs and those subscribed to the Mobile and Roam plans use the local reference frame (\texttt{FRAME\_UT}) based on the UT's local coordinate system.
Both the Mobile and Roam plans support mobility use cases.
Without a fixed reference system, the UT in motion must rely on its local alignment to determine beam steering angles.

In Fig.~\ref{fig:obstruction_maps}, we illustrate different obstruction maps from two UTs (UT A and UT B) of the same model (\texttt{rev3\_proto2}) placed in distinct obstructed environments.
Note that for the same UT hardware, the reference frame changes when the user switches to different subscription plans.
Fig.~\ref{fig:frame_earth} and Fig.~\ref{fig:frame_ut} were retrieved from the UT's gRPC interface when UT A was associated with the regular Residential plan and the Standby mode\footnote{The Standby mode is a new low-cost subscription plan offering unlimited data access at 500 Kbps, designed for emergency and backup purposes.}, respectively.
The structural obstruction pattern indicates the edge of a nearby building's rooftop.
UT A has a tilt angle of 26.5$^\circ$, and it is facing North with a boresight azimuth of about 3.2$^\circ$.
In \texttt{FRAME\_EARTH} obstruction maps (Fig.~\ref{fig:frame_earth}), the top-center pixel aligns with North, consistent with the top-down view from the Starlink mobile application (Fig.~\ref{fig:mobile_app}).
In contrast, in \texttt{FRAME\_UT} obstruction maps (Fig.~\ref{fig:frame_ut}), the bottom-center pixel always aligns with the UT's boresight direction, displaying the FOV from the UT's perspective.
For UT B in Fig.~\ref{fig:obstruction_maps}, it has a tilt angle of 7.7$^\circ$, and it also has the \texttt{FRAME\_UT} obstruction map reference frame type.
Metal shields were intentionally placed to the North of the UT, while the UT is facing North.
Consequently, the obstruction map in Fig.~\ref{fig:frame_ut_purpose_obstructed} indicates obstructions in the bottom-center region, aligning with the UT's boresight direction.
Additionally, as indicated by Fig.~\ref{fig:frame_ut} and Fig.~\ref{fig:frame_ut_purpose_obstructed}, the reference point~\cite{ahangarpourTrajectorybasedServingSatellite2024b}, i.e., the center pixel of the ellipse FOV in the \texttt{FRAME\_UT} obstruction map, along with the major and minor axes, changes in response to the tilt adjustments of the UT.

\section{Measurement Campaign}\label{sec:measurement}

In this section, we describe an overview of our mobile Starlink measurement setup and the data collection campaign.

\subsection{Methodology}\label{subsec:measurement-methodology}

\begin{figure}[tb]
    \centering
    \begin{subfigure}[h]{0.4\linewidth}
        \includegraphics[height=0.85\linewidth]{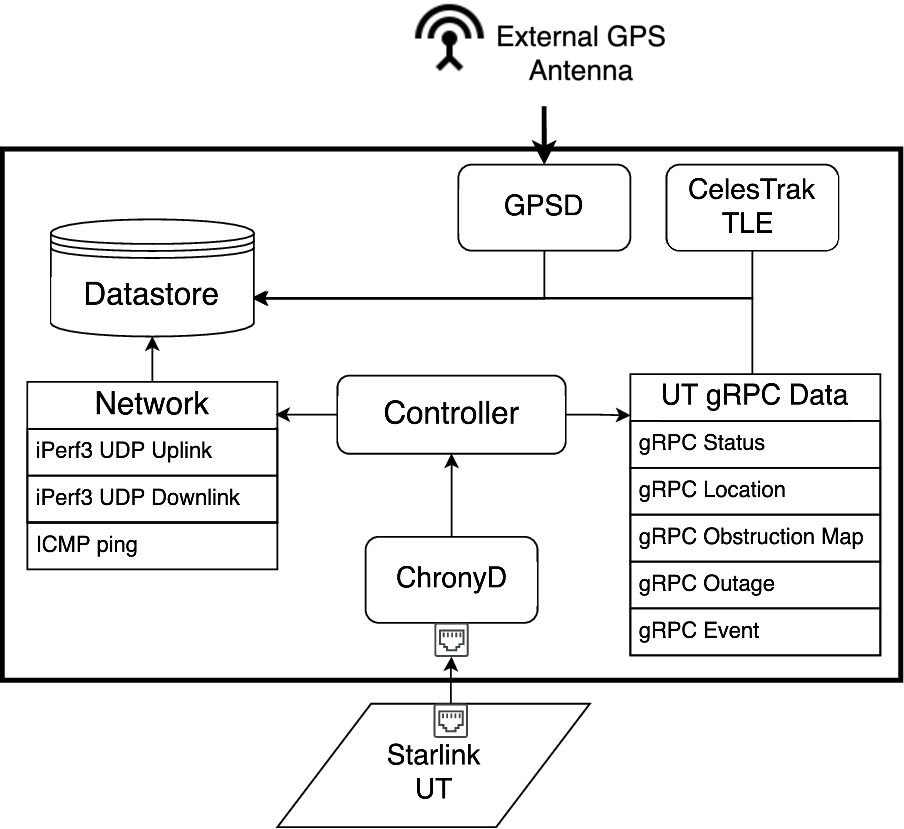}
        \caption{Data collection pipeline}
        \label{subfig:mobile-measurement-setup}
    \end{subfigure}
    \begin{subfigure}[h]{0.5\linewidth}
        \includegraphics[width=\linewidth]{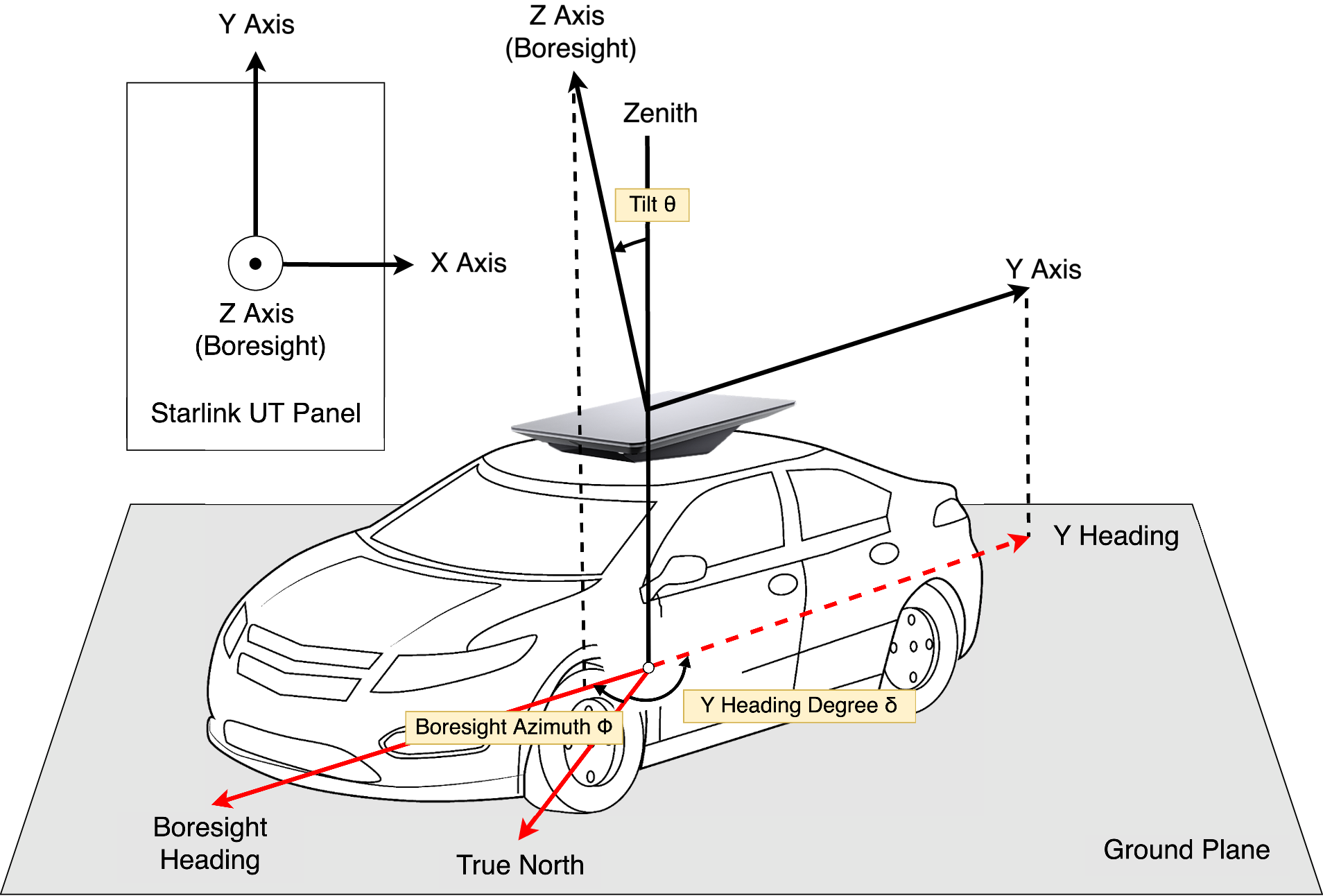}
        \caption{Mobile Starlink UT orientation}
        \label{fig:ut-orientation}
    \end{subfigure}
    \caption{Overview of measurement setup}
    \label{fig:leoviz-overview}
\end{figure}

Fig.~\ref{subfig:mobile-measurement-setup} presents an overview of our measurement setup
\footnote{The source code of our measurement toolkit will be made available at \url{https://github.com/clarkzjw/LEOViz}.}.
Our measurement setup consists of two main components: (i) Network performance metrics, and (ii) Starlink UT diagnostic information via the gRPC interface.

\noindent
\textbf{Latency:}
Starlink's default IPv4 Carrier-Grade NAT (CGNAT) gateway at \texttt{100}.\texttt{64}.\texttt{0}.\texttt{1} or the corresponding IPv6 gateways are reachable via ICMP-based \texttt{ping}, and are commonly used to measure the latency of satellite links in Starlink networks~\cite{panMeasuringSatelliteLinks2024b}.
The ICMP echo request interval is set to 10 ms.
Even though the actual packet interval may vary due to the scheduling behavior of the Linux operating system, however, by recording the ICMP response timestamps with the \texttt{-D} option, we can correlate the round-trip time (RTT) of ICMP requests with the regular 15-second handover intervals and dynamic beam switching events.

\noindent
\textbf{Throughput:}
We use \texttt{iPerf3} with UDP flows to measure the impact of regular handover and dynamic beam switching events on the throughput performance.
Note that we are not aiming to saturate the Starlink link capacity to measure the maximum achievable throughput, but rather to observe the throughput variations during handover and beam switching events.
Consequently, we set modest UDP target bitrates to 250 Mbps downlink on a Starlink flat panel high performance (FHP) UT.
UDP flows allow us to observe the throughput variations without being affected by different TCP congestion control algorithms.

\noindent
\textbf{Obstruction Map:}
We utilize the \texttt{starlink-grpc-tools}\footnote{\url{https://github.com/sparky8512/starlink-grpc-tools}} package and \texttt{grpcurl}\footnote{\url{https://github.com/fullstorydev/grpcurl}} to access the gRPC interface programmatically.
Note that the gRPC interface does not guarantee deterministic real-time behavior, and response times may vary.
In our measurements, we retrieve obstruction maps every 0.5 seconds, which avoids overloading the UT while ensuring we capture the most current snapshot of the obstruction map, as the obstruction map is updated about every second.

\noindent
\textbf{UT Alignment Status:}
The UT alignment status, as represented by the tilt angle and boresight azimuth, is obtained from the gRPC interface~\cite{ahangarpourTrajectorybasedServingSatellite2024b}.
Recent firmware updates also enabled the retrieval of dish quaternions, which we record for accurately determining the dish's orientation in \S\ref{sec:mobile-identification}.

\noindent
\textbf{Outage Events:}
Starlink UT can experience outages due to various reasons, such as obstructions (\texttt{OBSTRUCTED}), failure to receive satellite ephemeris from the central controller (\texttt{NO\_SCHEDULE}), unable to establish a downlink due to degraded signal quality (\texttt{NO\_DOWNLINK}) or unable to conduct a ping test to the PoP (\texttt{NO\_PINGS}), or is conducting a sky search before establishing a link (\texttt{SKY\_SEARCH}).
The response from the gRPC \texttt{get\_history} method includes a list of outage events, each specifying the event's start timestamp and duration in nanoseconds, and a boolean indicator of whether a dynamic beam switching event occurred.
A comprehensive list of Starlink UT event reasons is summarized in Table~\ref{tab:ut-outage-reasons}.

\noindent
\textbf{Location Information:}
The gRPC \texttt{get\_location} method returns the UT's positioning information as latitude and longitude using its built-in GNSS module.
An external GPS antenna, mounted alongside the UT and is connected to the measurement computer, enables the retrieval of detailed motion vectors from GPSD logs when the UT and the vehicle are in motion.

\noindent
\textbf{Timing Information:}
Each Starlink UT functions as a Stratum 1 NTP server at \texttt{192}.\texttt{168}.\texttt{100}.\texttt{1}, allowing our time-synchronized measurements to align with the regular handover intervals at 12-27-42-57 seconds of every minute.
However, it is important to note that the timing accuracy of asynchronous operations, including the response time of various gRPC calls on UT operations such as resetting the obstruction map, lacks strict real-time guarantees.
Consequently, the accuracy of timing information and satellite identification result is provided on a best-effort basis.

\subsection{Data Collection}

We begin with a controlled experimental phase using a UT mounted on a tripod, which allows us to validate our mobility-aware satellite identification method in isolation from vehicle motion and environmental variability. Using this setup, we verify the correctness of our orientation compensation and dynamic beam-switch detection under known, repeatable conditions.

We collected vehicular data covering approximately 500 km of driving over 5 hours across rural, urban, suburban, and highway environments in the midwestern USA. For these measurements, the FHP UT (\texttt{hp1\_proto2}) was mounted on the vehicle rooftop with a fixed 7.9\textdegree{} tilt, as illustrated in Fig.~\ref{fig:ut-orientation}.
As a stationary reference, we also performed the same measurements using a residential UT deployed under real-world obstruction condition as shown in Fig.~\ref{subfig:stationary-obstructed}.
Note that the stationary UT model (\texttt{rev3\_proto2}) has a smaller 110\textdegree{} FOV than FHP UT, and is used solely to validate identification accuracy under static obstructions rather than for throughput comparison.
In this paper, we focus on identifying communicating satellites in mobility scenarios and under different obstruction conditions, utilizing the diagnostic information from the same UT to explain network performance variations, without making direct cross-comparisons between different dish models on network performance.

\section{Mobility-Aware Starlink Satellites Identification}\label{sec:mobile-identification}

In this section, we first present the challenges of accurately identifying communicating satellites in mobility scenarios.
Then, we discuss the dynamic beam switching behavior when the UT faces transient obstructions while in motion.
Finally, we propose a mobility-aware Starlink satellite identification mechanism.

\subsection{Accurate Determination of Dish Orientation}\label{subsec:dish_orientation}

In Fig.~\ref{fig:ut-orientation}, we illustrated the mobile Starlink UT's orientation in 3D space, represented by two different reference systems.
The first representation contains its tilt angle $\theta$ and the boresight azimuth $\phi$, which depict the UT's orientation in Euclidean space, relative to the ground plane~\cite{ahangarpourTrajectorybasedServingSatellite2024b}.
Additionally, the UT's orientation can also be represented using quaternions, which is provided by the gRPC interface as \texttt{ned2dishQuaternion}.
Quaternion is a four-dimensional representation using complex numbers, widely employed in fields such as robotics and 3D computer graphics for handling object rotations.
They are preferred over other methods, such as Euler angles, because they avoid issues like ``gimbal lock''.
The \texttt{ned2dishQuaternion} metric exposed from the gRPC interface represents a unit quaternion $\mathbf{q}$, with \texttt{qScalar} ($w$), \texttt{qX} ($x$), \texttt{qY} ($y$), and \texttt{qZ} ($z$), where $\mathbf{q} = w + x \mathbf{i} + y \mathbf{j} + z \mathbf{k}$, and $\left\Vert\mathbf{q}\right\Vert = \sqrt{w^2 + x^2 + y^2 + z^2} = 1$.
$\mathbf{i}$, $\mathbf{j}$, and $\mathbf{k}$ are imaginary units, where $\mathbf{i}^2 = \mathbf{j}^2 = \mathbf{k}^2 = -1$,
$\mathbf{i}\mathbf{j}= \mathbf{k}$, $\mathbf{j}\mathbf{k}= \mathbf{i}$, and $\mathbf{k}\mathbf{i}= \mathbf{j}$.

When the UT is mounted on vehicles with a fixed tilt angle $\theta$ and facing the vehicle's forward direction, as shown in Fig.~\ref{fig:ut-orientation}, the $+Z$ axis is oriented upwards towards the boresight, with antenna elements on the UT panel facing the sky.
The $+Y$ axis extends from the center to the top of the UT panel, and the $+X$ axis, perpendicular to the $+Y$ axis, points to the right in the figure in accordance with the right-handed coordinate system.
Consequently, the boresight heading, which is the projection of the $+Z$ axis onto the ground plane, indicates the UT's forward direction.
Meanwhile, the projection of the $+Y$ axis points in the opposite direction to the boresight heading, with a 180\textdegree{} offset.
We define the angle of this projection, measured from the True North, in the direction opposite to the boresight azimuth, as the ``Y heading degree'', denoted by $\delta$.
The unit quaternion $\mathbf{q}$ can be converted to $\delta$ using the Tait-Bryan angles conversion for yaw, pitch, and roll-based representations~\cite{jackQuaternionsRotationSequences2002}.

\begin{figure}[tb]
    \centering
    \includegraphics[width=0.5\linewidth]{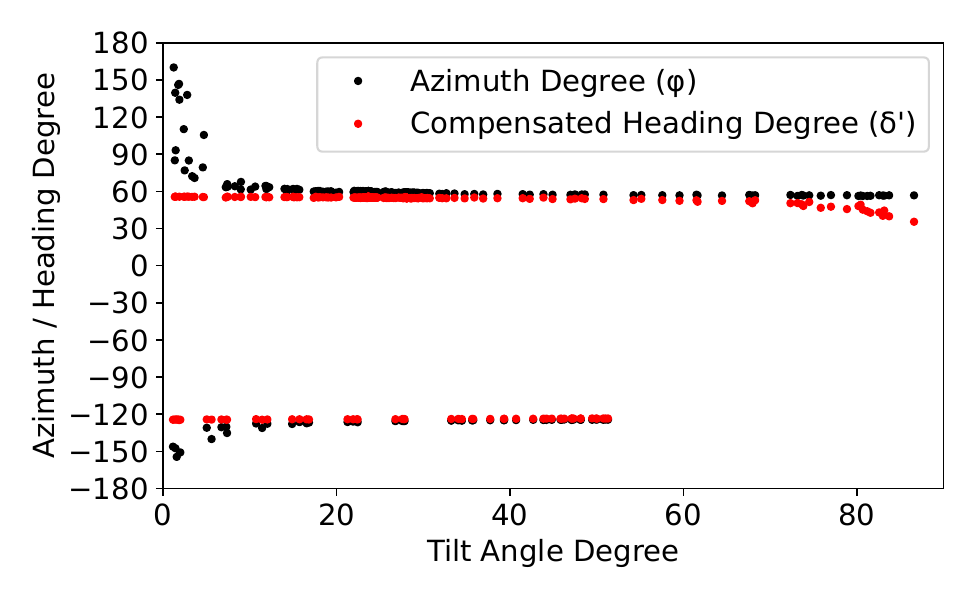}
    \caption{The compensated heading degree $\delta^\prime$ converted from $\mathbf{q}$ remains stable when the UT is near level ($\theta \xrightarrow{} 0$)}
    \label{fig:inaccurate-azimuth}
\end{figure}

In certain extreme scenarios, such as when the UT is mounted on marine vessels affected by ocean waves, the UT might temporarily face backward relative to the vessel's forward direction.
In practice, varying road conditions may often cause the UT panel to reach a near-level position.
When the tilt angle $\theta \xrightarrow{} 0$, accurately determining boresight azimuth $\phi$ becomes challenging, as the projection of the $+Z$ axis onto the ground plane becomes poorly defined.
With the UT panel being almost flat, slight tilt or orientation changes can disproportionately impact the accuracy of boresight azimuth $\phi$, because the system is more sensitive to angular variations.

To demonstrate the impact of inaccurate boresight azimuth measurements, we mounted the UT on a tripod and manually oriented it towards $\phi^\prime \approx 60^\circ$.
We only adjust the tilt angle $\theta$, and record the UT's boresight azimuth degree $\phi$, as reported by the gRPC interface.
Note that when the UT rotates along its $X$ axis and faces the direction opposite to the vehicle's forward movement, the gRPC interface still reports positive $\theta$ values.
However, the boresight azimuth $\phi$ now includes a 180\textdegree{} offset to the vehicle's forward direction, as illustrated by $\phi \approx -120^\circ$ in Fig.~\ref{fig:inaccurate-azimuth}.
As the tilt angle $\theta \xrightarrow{} 0$, boresight azimuth $\phi$ starts to become inaccurate when $\theta < 15^\circ$, which is the typical tilt angle range when mounting a flat panel UT on vehicles.
As shown in Fig.~\ref{fig:inaccurate-azimuth}, by converting the quaternion $\mathbf{q}$, we can calculate a more robust alternative to boresight azimuth by using the compensated heading degree $\delta^\prime=\delta+180^\circ$, when the pitch angle along the $+Y$ axis is positive.
$\delta^\prime$ remains stable across different $\theta$ values, since it stays orthogonal to the boresight, thereby preventing inaccurate projections.
This remains true unless $\theta \xrightarrow{} 90^\circ$, a situation unlikely to occur in realistic scenarios.

\begin{figure}[tb]
    \centering
    \begin{subfigure}[h]{0.14\linewidth}
        \includegraphics[width=\linewidth]{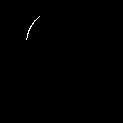}
        \caption{$f_{M_{t-1}}$}
    \end{subfigure}
    \begin{subfigure}[h]{0.14\linewidth}
        \includegraphics[width=\linewidth]{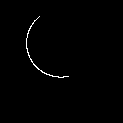}
        \caption{$f_{M_t}$}
    \end{subfigure}
    \begin{subfigure}[h]{0.14\linewidth}
        \includegraphics[width=\linewidth]{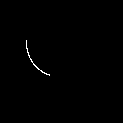}
        \caption{$f_{M_{t-1}} \oplus f_{M_{t}}$}
        \label{subfig:xor_frame}
    \end{subfigure}
    \caption{When the UT is in motion, the $\oplus$ operation can return multiple pixels between obstruction maps}
    \label{fig:obstruction-map-fixed-rotation}
    \vspace{-1em}
\end{figure}

\subsection{Impact of UT Mobility on Obstruction Maps}

To investigate how vehicular movements and UT mobility affect the satellite trajectories in obstruction maps, we conducted controlled rotation tests.
We mounted the UT on a tripod at a fixed location with a fixed tilt angle $\theta$, and rotated only the tripod head to orient the UT in different directions, thereby changing the boresight azimuth $\phi$.
As shown in Fig.~\ref{fig:obstruction-map-fixed-rotation}, as the UT rotates, the observed satellite trajectories in the obstruction map stretch accordingly, forming a continuous arc whose curve length varies with the rotation speed, while its diameter depends on the elevation of the communicating satellites.
This indicates that when the UT is in motion, its beam steering capability enables it to continuously track communicating satellites as long as they remain within the FOV, depending on the elevation of communicating satellites.
Previous studies~\cite{tanveerMakingSenseConstellations2023c,ahangarpourTrajectorybasedServingSatellite2024b,liuVivisectingStarlinkThroughput2025} use the XOR ($\oplus$) operation between adjacent obstruction maps to determine the pixel coordinates of the current communicating satellites in stationary scenarios.
When the UT is in motion, the $\oplus$ operation usually yield a line segment between adjacent frames, rather than only a few pixels, as shown in Fig.~\ref{subfig:xor_frame}.
We address this challenge in \S\ref{subsec:mobility-identification}.

\begin{figure}[tb]
    \centering
    \begin{subfigure}[h]{0.15\linewidth}
        \includegraphics[width=\linewidth]{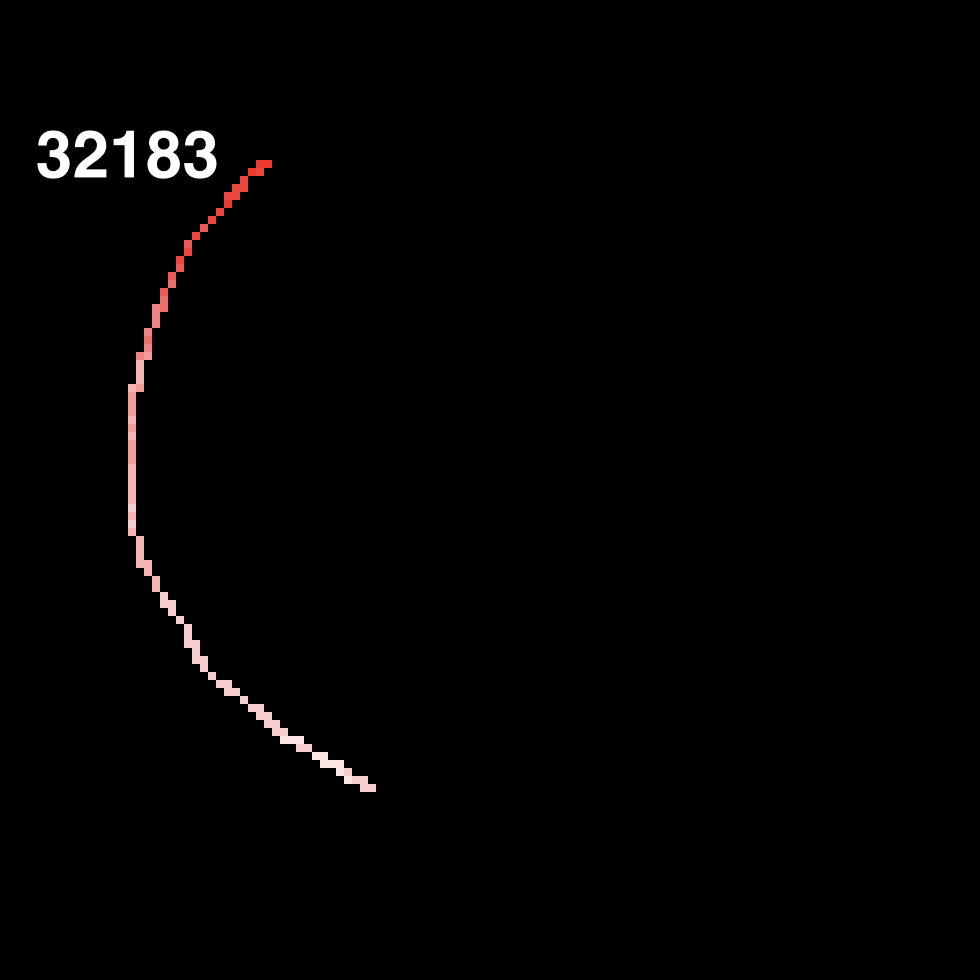}
        \caption{\texttt{12}:\texttt{39}.\texttt{912}}
        \label{subfig:obstruction-1}
    \end{subfigure}
    \begin{subfigure}[h]{0.15\linewidth}
        \includegraphics[width=\linewidth]{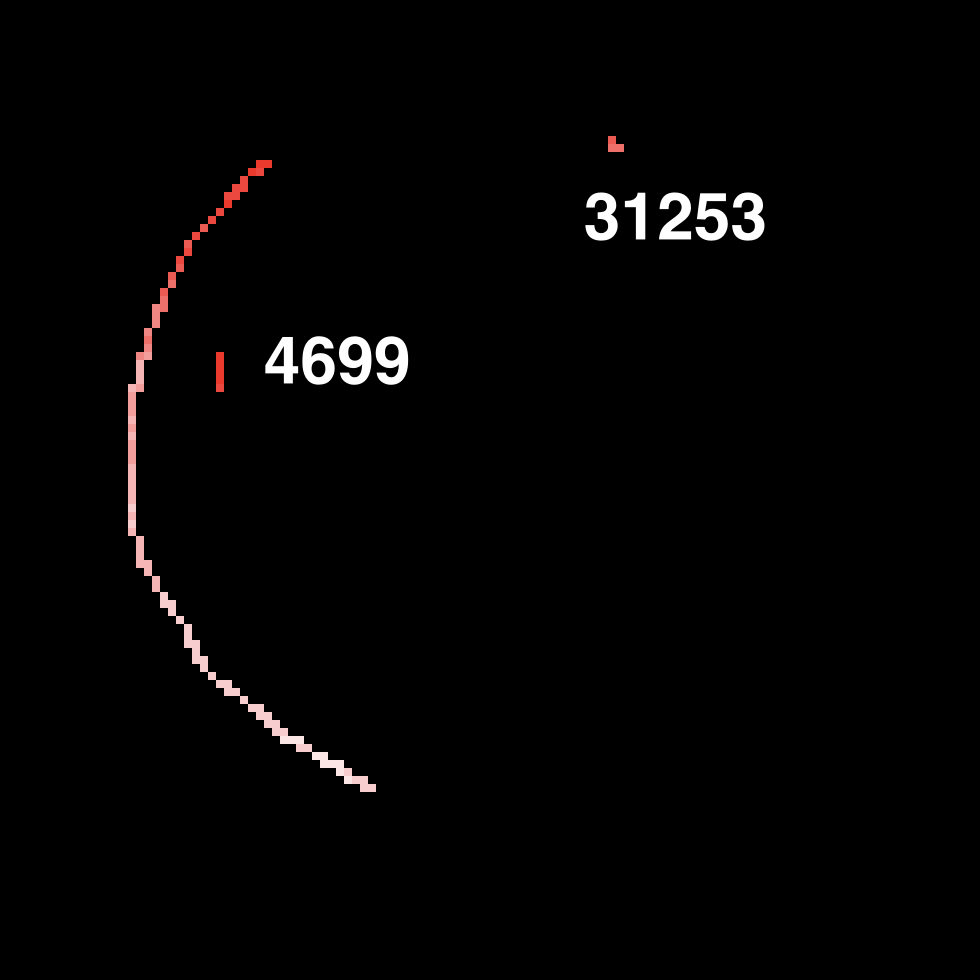}
        \caption{\texttt{12}:\texttt{40}.\texttt{463}}
        \label{subfig:obstruction-2}
    \end{subfigure}
    \begin{subfigure}[h]{0.15\linewidth}
        \includegraphics[width=\linewidth]{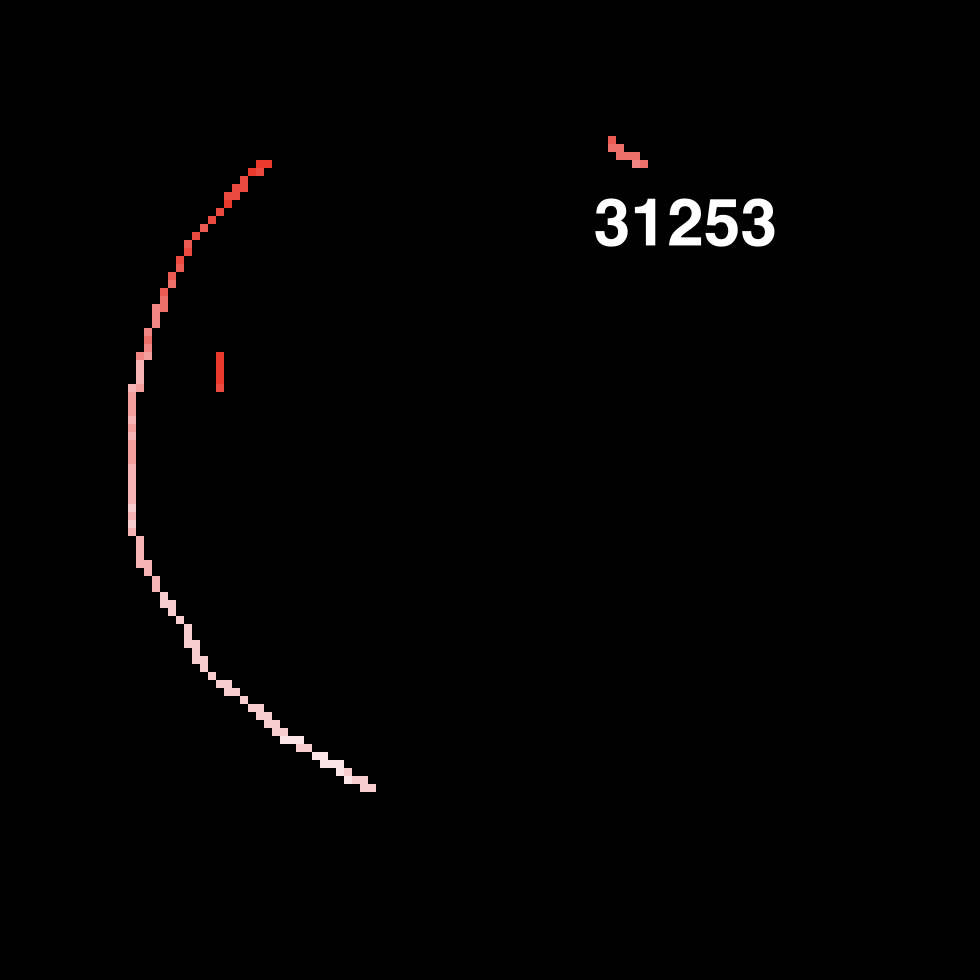}
        \caption{\texttt{12}:\texttt{41}.\texttt{058}}
        \label{subfig:obstruction-3}
    \end{subfigure}
    \begin{subfigure}[h]{0.15\linewidth}
        \includegraphics[width=\linewidth]{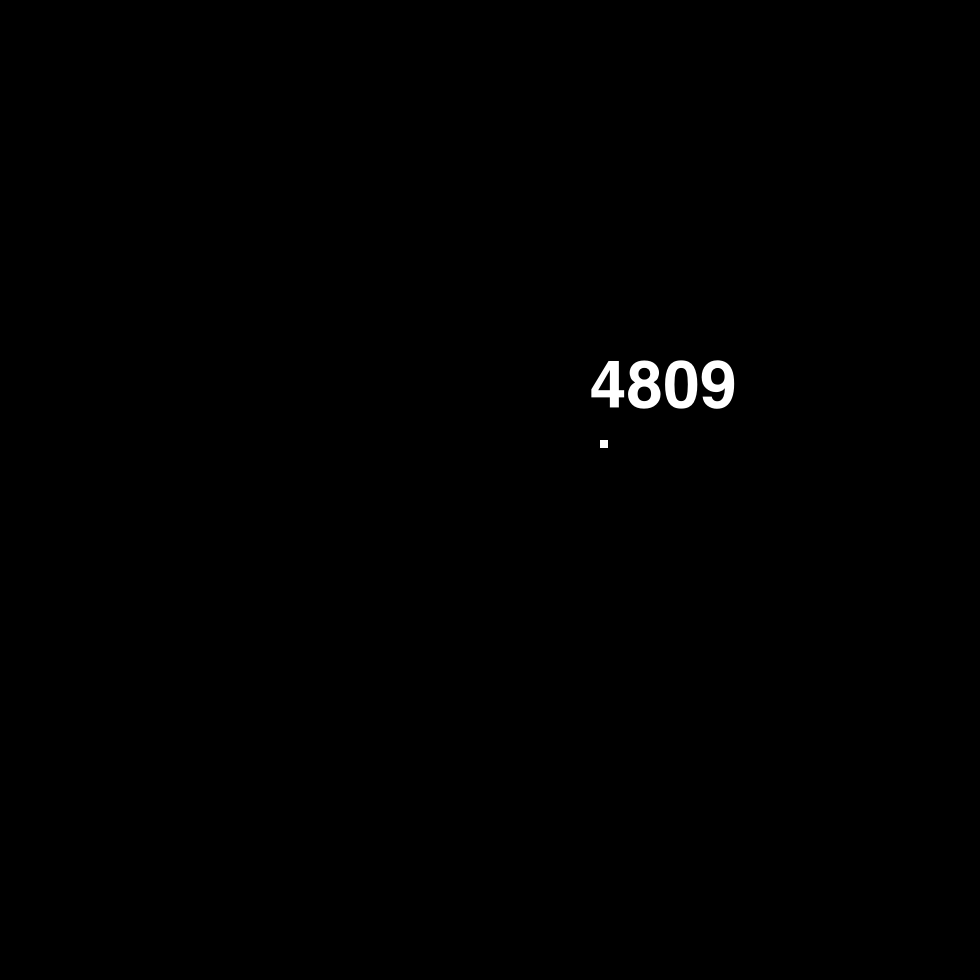}
        \caption{\texttt{12}:\texttt{42}.\texttt{338}}
        \label{subfig:obstruction-4}
    \end{subfigure}
    \begin{subfigure}[h]{0.15\linewidth}
        \includegraphics[width=\linewidth]{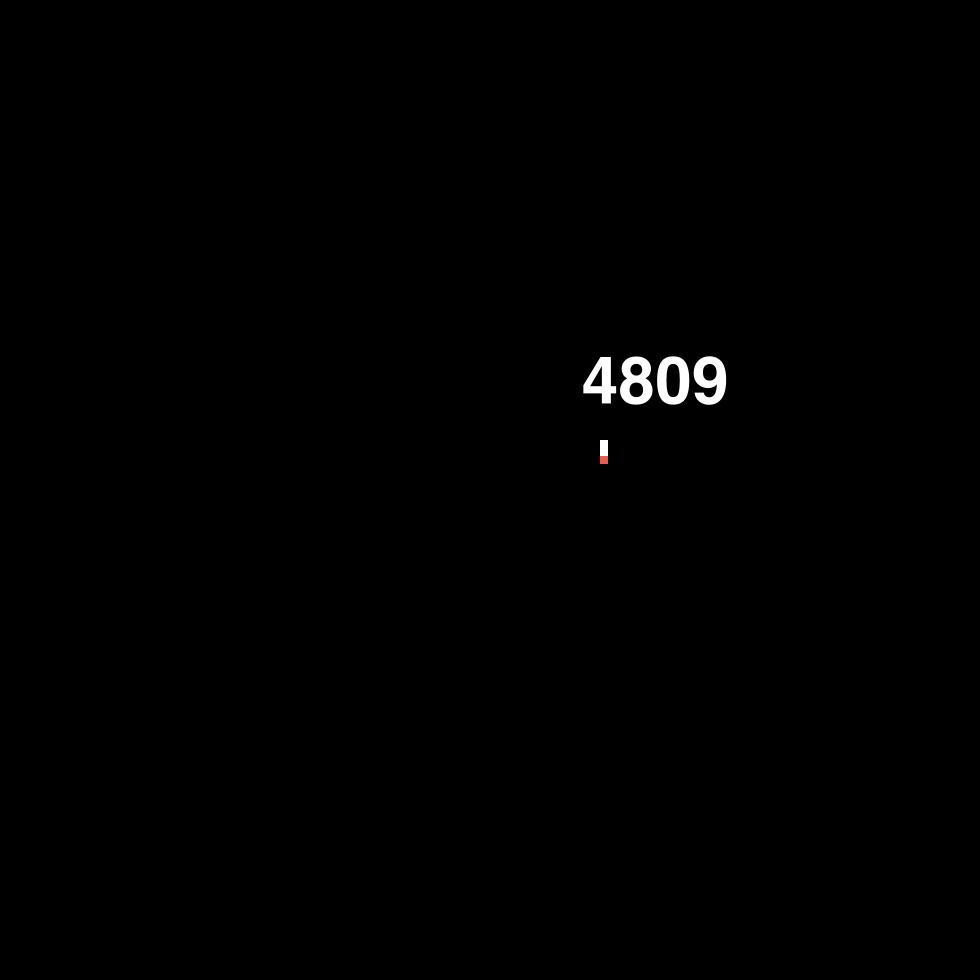}
        \caption{\texttt{12}:\texttt{44}.\texttt{833}}
        \label{subfig:obstruction-5}
    \end{subfigure}
    \begin{subfigure}[h]{0.15\linewidth}
        \includegraphics[width=\linewidth]{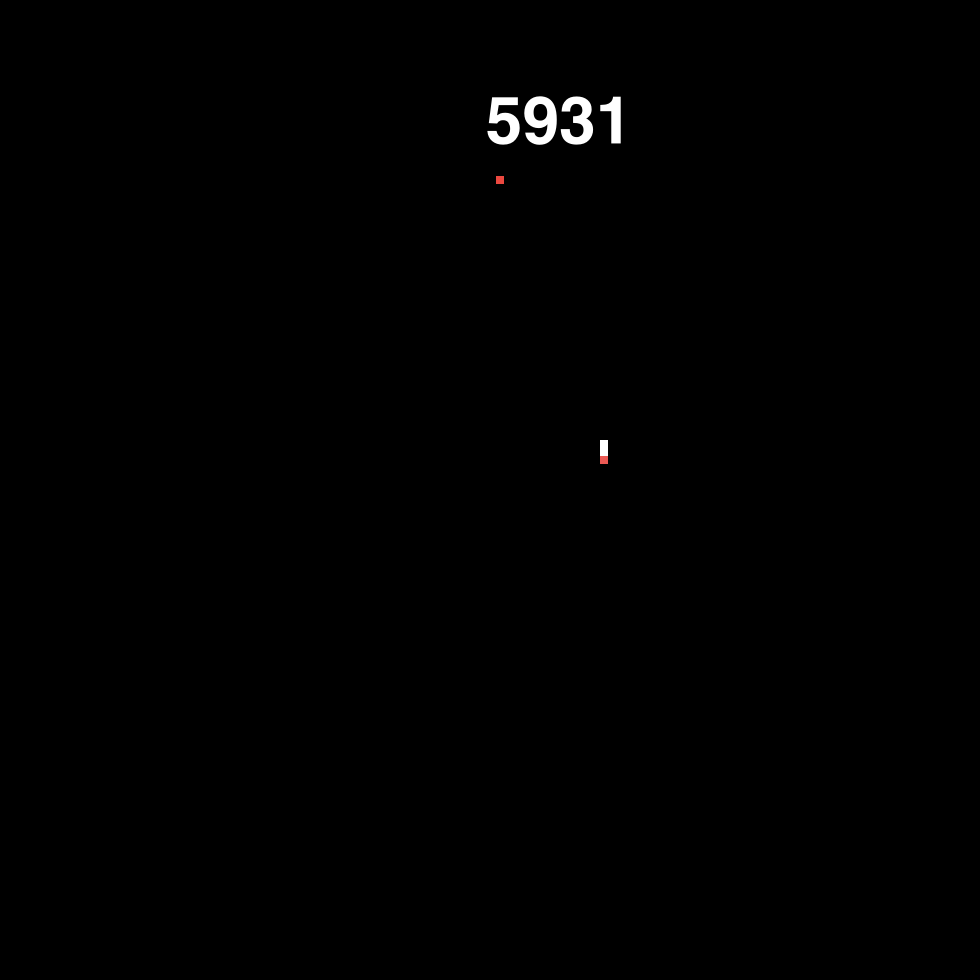}
        \caption{\texttt{12}:\texttt{45}.\texttt{427}}
        \label{subfig:obstruction-6}
    \end{subfigure}
    \begin{subfigure}[h]{\linewidth}
        \centering
        \includegraphics[width=0.6\linewidth]{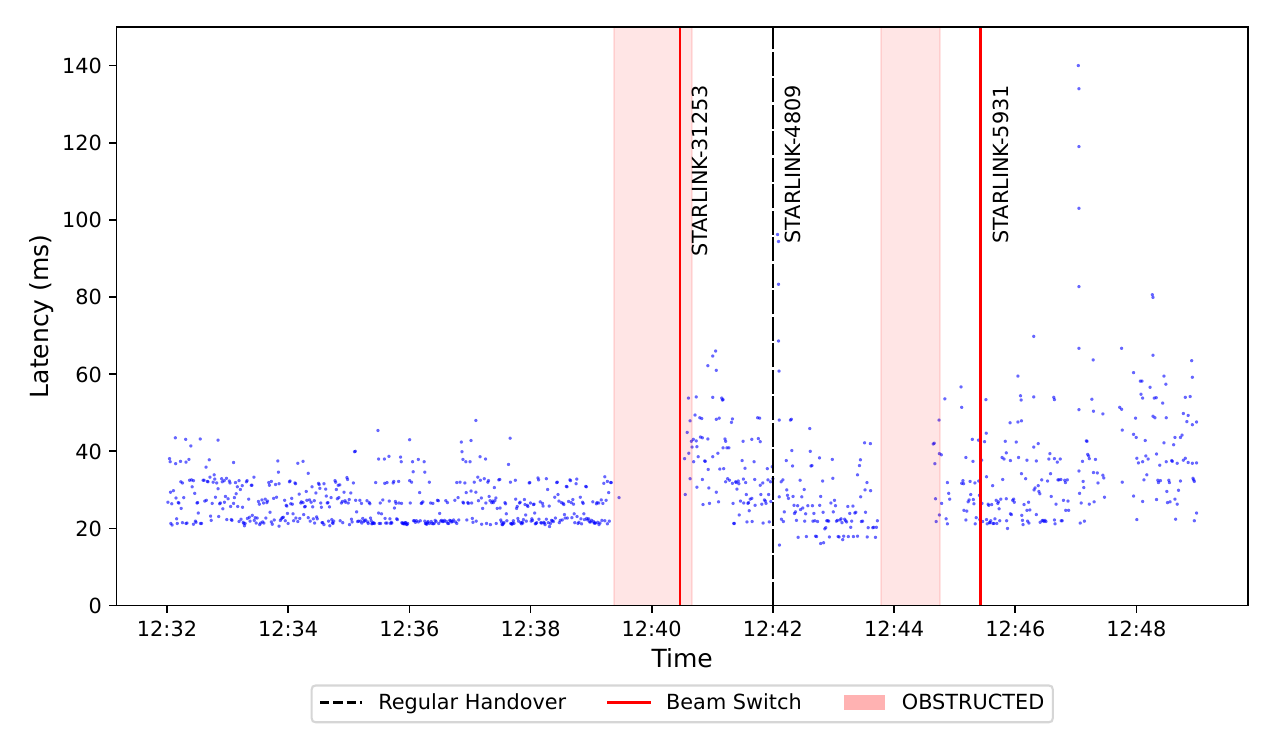}
        \caption{Time synchronized ICMP \texttt{ping} latency test with regular handover, beam switching, and outage events}
        \label{subfig:latency}
    \end{subfigure}
    \caption{A Starlink UT in motion conducts multiple reactive beam switching attempts, while unpredictable transient obstructions on the road continue to cause network interruptions}
    \label{fig:beam-fast-switching}
    \vspace{-1em}
\end{figure}

\subsection{Analysing Dynamic Beam Switching Behavior}

Existing Starlink satellite identification algorithm~\cite{tanveerMakingSenseConstellations2023c,ahangarpourTrajectorybasedServingSatellite2024b,liuVivisectingStarlinkThroughput2025} assumed that a UT stays connected to the same satellite throughout the entire 15-second handover timeslot.
This assumption simplifies the satellite identification process in stationary scenarios.
For each 15-second timeslot, the trajectories from this timeslot are compared with all candidate satellites within the UT's FOV to identify the best matching satellite using custom distance metrics.
This assumption generally holds true as dynamic beam switching events are rare in stationary and obstruction-free scenarios.
Thus, the trajectories from the same 15-second timeslot are typically continuous from a single satellite.
However, with dynamic beam switching events in obstructed or mobility scenarios, blindly fitting the entire trajectory from a timeslot to a single satellite produces inaccurate identification results.

To the best of our knowledge, our analysis is the first to show that when a UT is obstructed, dynamic beam switching events can be triggered, causing the UT to switch to different satellites, even within the same 15-second timeslot.
Moreover, dynamic beam switching events near the boundaries of regular handover intervals can exacerbate network performance degradation if the UT is subsequently handed over to another satellite.
In Fig.~\ref{subfig:obstruction-1}, the vehicle was making a U-turn in road sections with dense foliage coverage, which resulted in the trajectory of the connected satellite \texttt{STARLINK-32183} being stretched into an arc.
At around \texttt{12}:\texttt{40}.\texttt{463}, two new spatially distinct line segments appeared in the obstruction map in Fig.~\ref{subfig:obstruction-2}, corresponding to \texttt{STARLINK-4699} and \texttt{STARLINK-31253}.
We cannot determine the order of satellites that the UT attempted to connect with, solely based on the XOR operation between these two frames,
However, in Fig.~\ref{subfig:obstruction-3}, the trajectory of \texttt{STARLINK-31253} on the top right corner continues to grow, indicating the sequence of connections.
As the regular handover time approaches, despite the UT had recently conducted a dynamic beam switching, the dish was handed over to a new satellite \texttt{STARLINK-4809}, indicated by another spatially distinct pixel in Fig.~\ref{subfig:obstruction-4} at \texttt{12}:\texttt{42}.\texttt{338}.
After the regular handover, at around \texttt{12}:\texttt{44}.\texttt{833}, the UT encountered another transient obstruction on the road, as indicated by the single red pixel in Fig.~\ref{subfig:obstruction-5}.
As a result, another dynamic beam switching event occurred, switching the UT to \texttt{STARLINK-5931}, as shown in Fig.~\ref{subfig:obstruction-6}.

To understand how the series of dynamic beam switching events impact network performance, we synchronized the ICMP \texttt{ping} latency test results with the obstructions and outage events, as shown in Fig.~\ref{subfig:latency}.
The outage events exported from the gRPC interface show that an \texttt{OBSTRUCTED} event occurred at approximately \texttt{12}:\texttt{39}.\texttt{380} and lasted for 1,280,066,127 nanoseconds, or about 1.28 seconds, ending at \texttt{12}:\texttt{40}.\texttt{660}.
Meanwhile, for the two dynamic beam switching attempts that happened in Fig.~\ref{subfig:obstruction-2}, UT only reported one single outage event.
This is likely due to the network link experienced continuous outages throughout this period, as indicated by the RTT gap shown in Fig.~\ref{subfig:latency}.
Note that we identified \texttt{STARLINK-31253} before the first outage event concluded, due to the presence of obstructed pixels in Fig.~\ref{subfig:obstruction-2}.
In contrast, because of the gRPC interface lacks strict real-time guarantee, \texttt{STARLINK-5931} was identified only after the outage ended.

Consequently, due to the dynamic beam switching behavior, the existing Starlink satellite identification algorithms~\cite{tanveerMakingSenseConstellations2023c,ahangarpourTrajectorybasedServingSatellite2024b,liuVivisectingStarlinkThroughput2025} would produce inaccurate identification results, as these algorithms incorrectly use pixel coordinates from different satellite trajectories to fit a best match.
When the UT is in motion, transient obstructions are more likely to block the LoS between the UT and communicating satellites, thus triggering dynamic beam switching events more frequently.

\begin{figure*}[tb]
    \centering
    \includegraphics[width=0.9\linewidth]{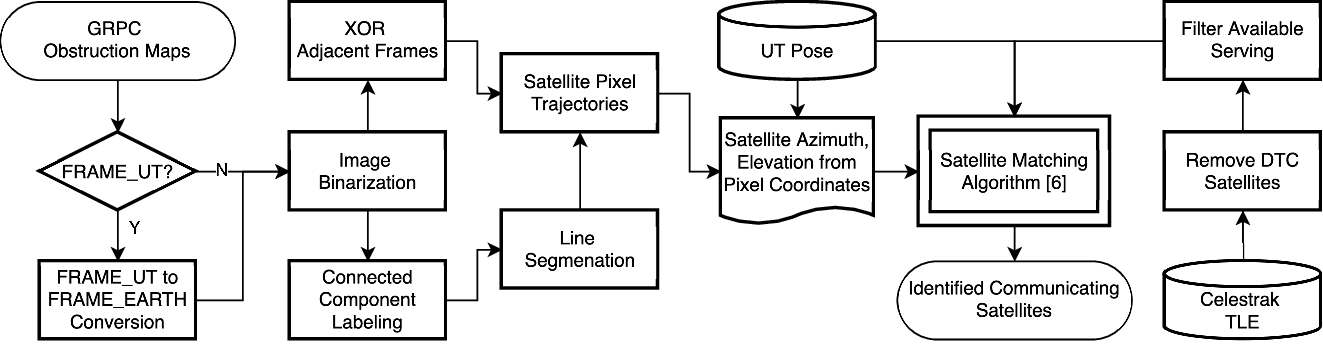}
    \caption{Flow chart of the mobility-aware satellite identification and beam switching detection algorithm}
    \label{fig:flow-chart}
    \vspace{-1em}
\end{figure*}

\subsection{Mobility-Aware Satellite Identification}\label{subsec:mobility-identification}

To overcome this challenge and accurately identify communicating satellites amid potential beam switching behavior, we propose a custom mobility-aware satellite identification algorithm with dynamic beam switching detection.
Fig.~\ref{fig:flow-chart} shows the overall flow chart of the algorithm.
For mobile UTs, the obstruction maps retrieved from the gRPC interface are represented in the \texttt{FRAME\_UT} projection.
For consistency and a unified workflow, we first convert the pixel coordinates in the \texttt{FRAME\_UT} representation to \texttt{FRAME\_EARTH} representation by compensating the UT's orientation.
Then, image binarization is applied to the processed \texttt{FRAME\_EARTH} obstruction maps based on the presence of pixels.
For each binarized obstruction map frame $f_{M_t}$, we compute $f_{M_t\oplus}$$=$$f_{M_{t-1}} \oplus f_{M_{t}}$.
When the UT is in motion, $f_{M_t\oplus}$ typically contains line segments whose lengths vary with vehicle speed, and whose shapes change with vehicle heading adjustments.

A dynamic beam switching event results in a new, spatially distinct line segment.
We detect such changes using connected component labeling (CCL) algorithms~\cite{sametEfficientComponentLabeling1988} on binarized obstruction maps.
We sample the midpoint of each line segment in $f_{M_t\oplus}$ as the projected pixel coordinates of the communicating satellites when capturing $f_{M_t}$.
When $f_{M_t\oplus}$ contains multiple line segments such as in the case of Fig.~\ref{subfig:obstruction-2}, we utilize the next frame $f_{M_{t+1}}$ to identify the sequence of connected satellites.
Among successive XOR frames $f_{M_t\oplus}$ and $f_{M_{t+1}\oplus}$, if the number of line segments increases, we process each detected line segment, i.e., the trajectory of communicating satellites, independently.
By converting the pixel coordinates from obstruction maps to the celestial coordinates of azimuth and elevation for communicating satellites, and comparing these with the visible satellites within the UT's FOV, considering the UT's location and orientation with the compensated heading $\delta^\prime$ from \S\ref{subsec:dish_orientation}, we then apply the satellite matching algorithm from~\cite{ahangarpourTrajectorybasedServingSatellite2024b} to identify the best matching satellite for every line segment.
As a result, multiple communicating satellites can be identified within the same 15-second timeslot, if dynamic beam switching events happened.
It is also important to note that the Starlink satellite ephemeris data on CelesTrak includes satellites dedicated for direct-to-cell (DTC) services.
Current DTC satellites do not use broadband payloads.
As a result, we exclude them when identifying communicating satellites.

\section{Network Performance Measurements under Beam Switching}\label{sec:analysis}

In this section, we first present a detailed analysis and case studies on how obstructions lead to dynamic beam switching events, significantly impacting network performance in both stationary and mobile scenarios.
Then, we illustrate how UT mobility influences the FOV and the visible satellites, contributing to more frequent beam switching events in mobile UTs.
Finally, we assess the accuracy of the proposed mobility-aware satellite identification algorithm.

Table~\ref{table:outage-event-statistics} summarizes the outage event statistics observed in both mobile and stationary obstructed scenarios.
Both statistics are collected over a 5-hour measurement period.
Note that the stationary obstructed UT (Fig.~\ref{subfig:stationary-obstructed}) experienced more \texttt{OBSTRUCTED} outages due to the dense tree coverage.
However, the mobile UT encounted more \texttt{SKY\_SEARCH} outages, as transient obstructions such as highway overpasses frequently blocked the LoS during movement or FOV changes due to UT mobility, and it took longer for the UT to establish a new connection with a visible satellite.
Additionally, long term performance monitoring indicates that, for the UT in Fig.~\ref{subfig:stationary-obstructed}, the average obstructed time ratio has been reduced from around 10\% in 2024 to less than 1\% in December 2025, contributing to the proactive beam switching mechanism and the increased satellite density.

\begin{figure}[tb]
    \centering
    \begin{subfigure}[h]{\linewidth}
        \centering
        \includegraphics[width=0.55\linewidth]{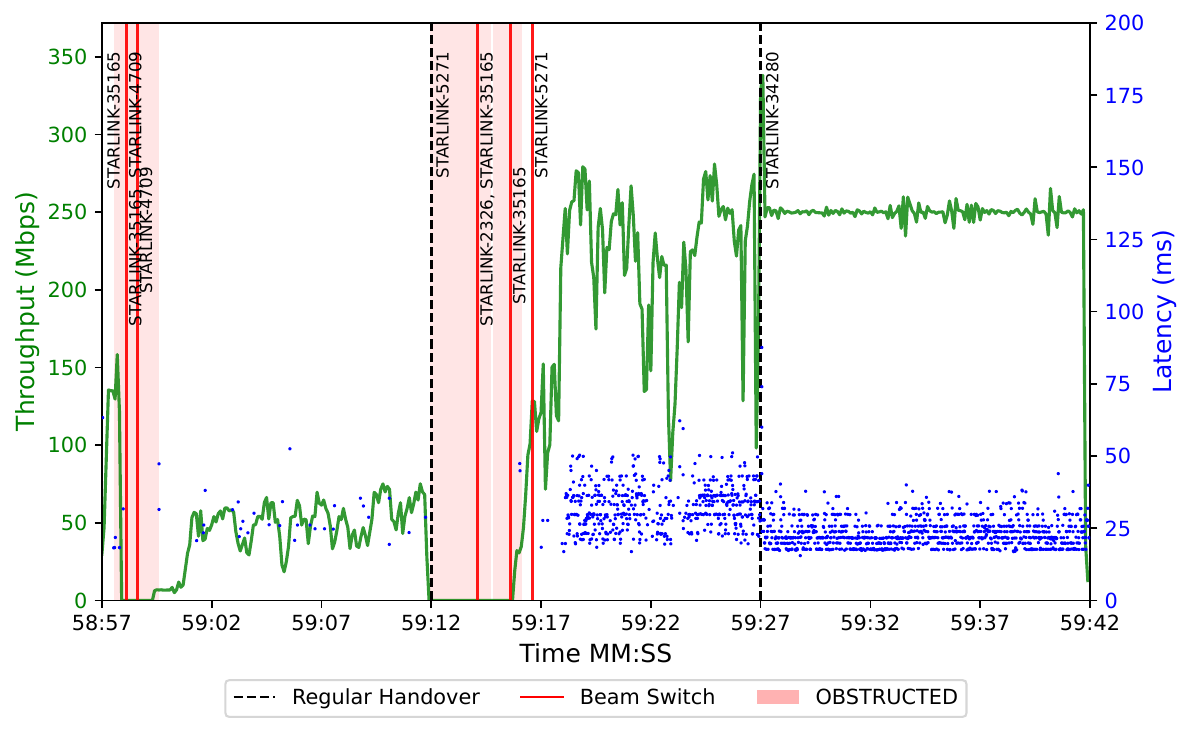}
        \caption{Multiple timeslots of outage events and network performance on a stationary UT with partial obstructions}
        \label{subfig:victoria-seattle-rtt}
    \end{subfigure}
    \begin{subfigure}[h]{0.15\linewidth}
        \centering
        \includegraphics[width=\linewidth]{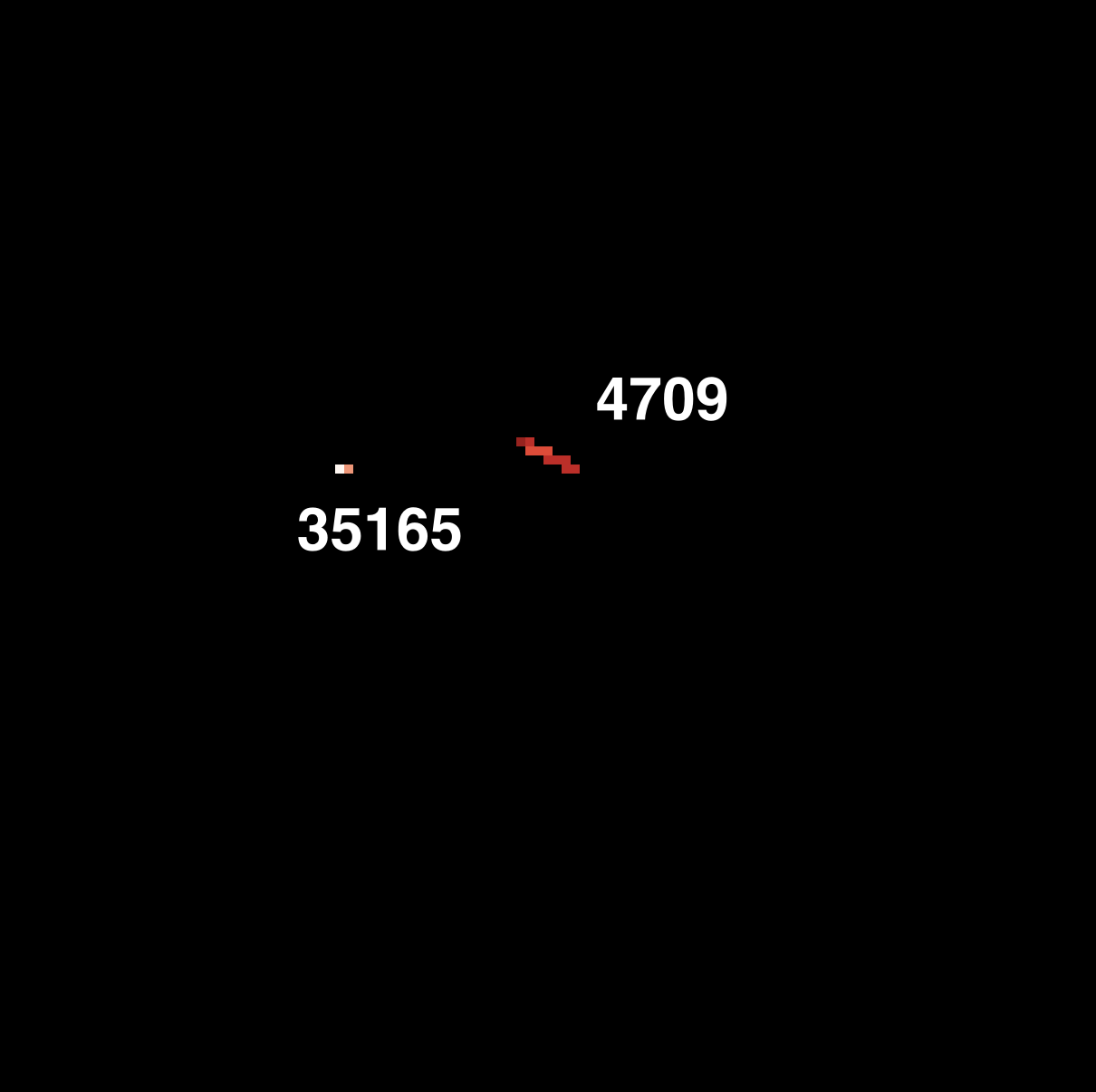}
        \caption{Timeslot 1}
        \label{subfig:s2_t1}
    \end{subfigure}
    \begin{subfigure}[h]{0.15\linewidth}
        \centering
        \includegraphics[width=\linewidth]{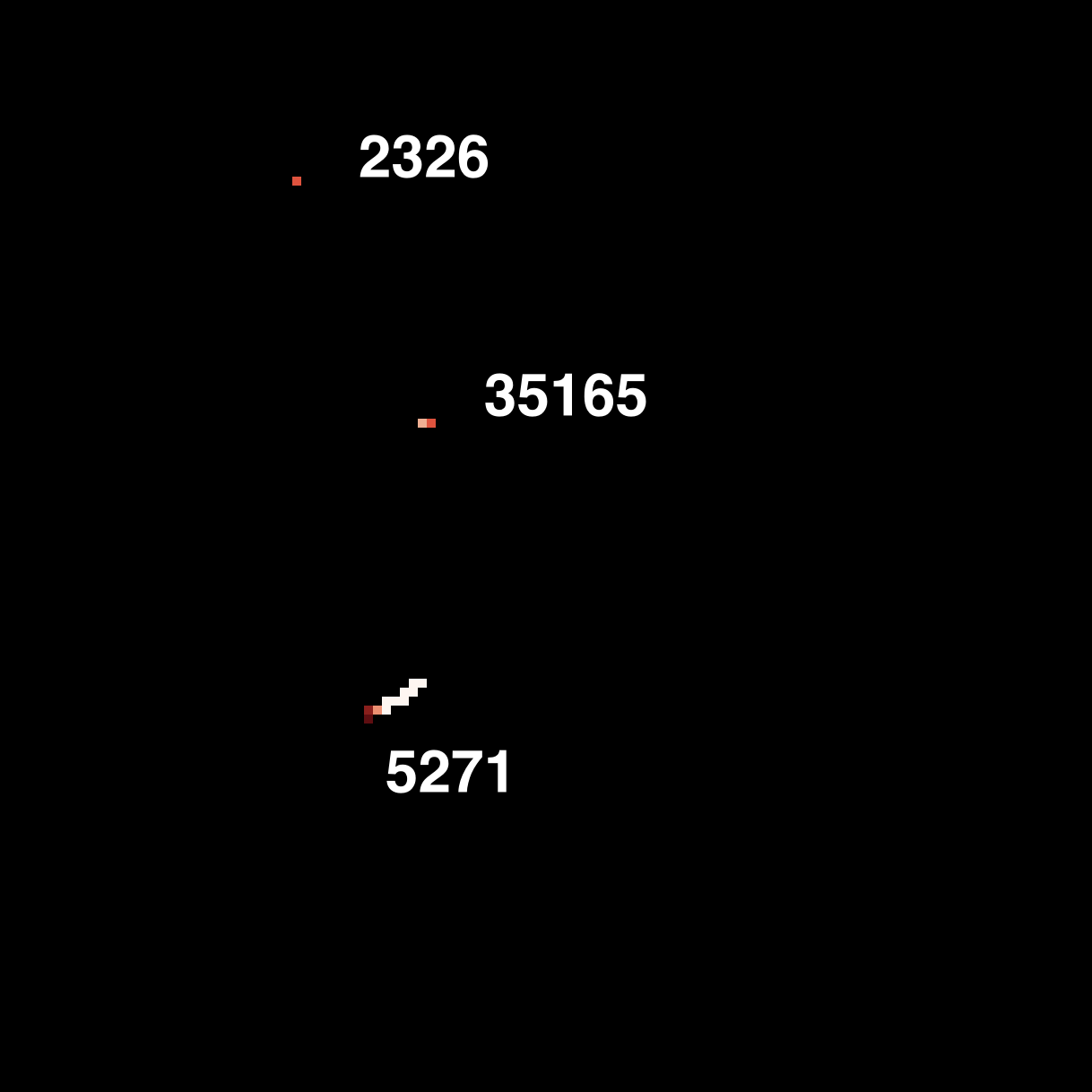}
        \caption{Timeslot 2}
        \label{subfig:s2_t2}
    \end{subfigure}
    \begin{subfigure}[h]{0.15\linewidth}
        \centering
        \includegraphics[width=\linewidth]{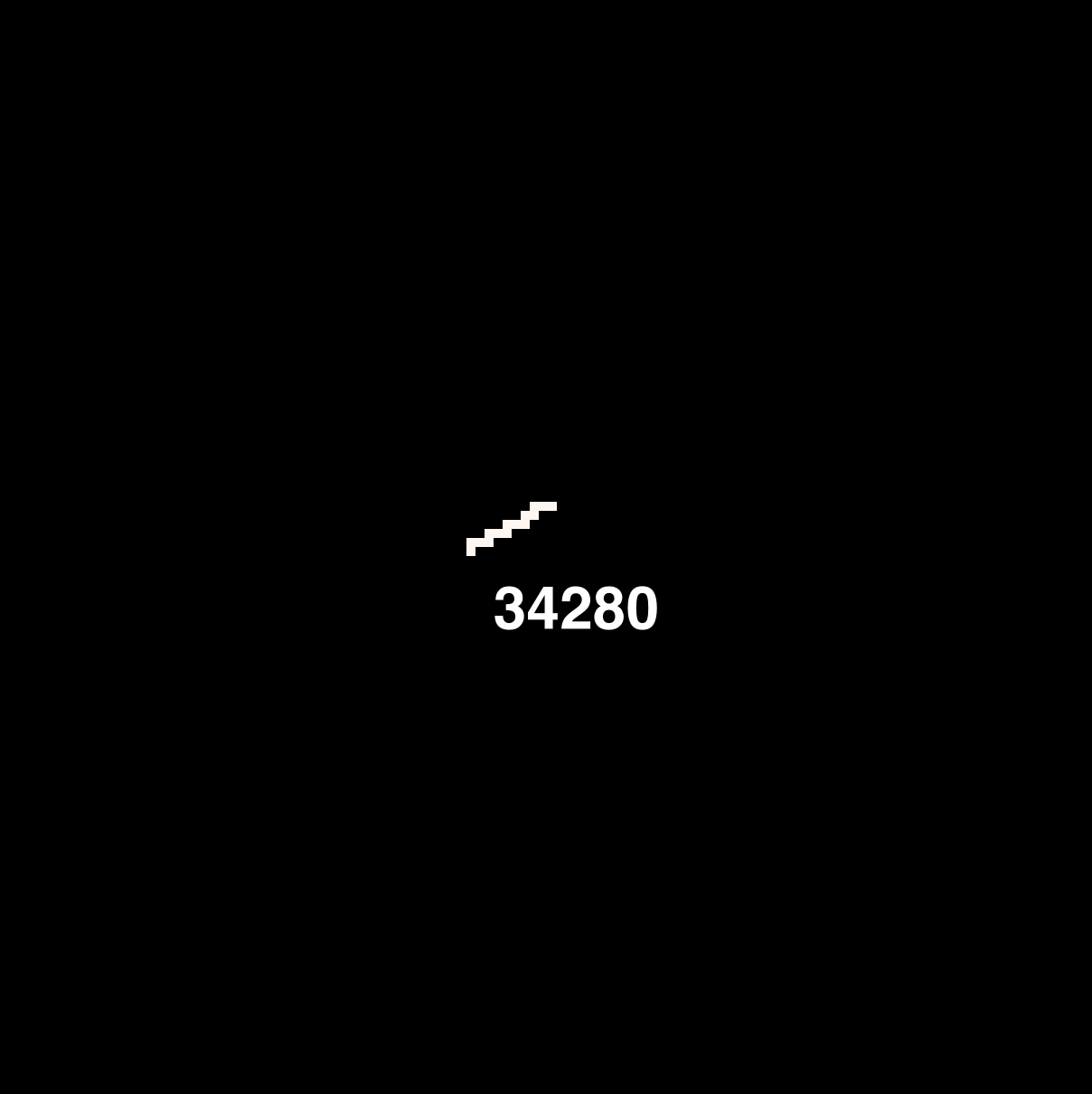}
        \caption{Timeslot 3}
        \label{subfig:s2_t3}
    \end{subfigure}
    \caption{A stationary UT conducts dynamic beam switching to circumvent obstructions. However, multiple failed switching attempts occurred, causing prolonged network interruptions.}
    \label{fig:latency-stationary-compare}
    \vspace{-1.2em}
\end{figure}

\subsection{Stationary UT}

Fig.~\ref{subfig:victoria-seattle-rtt} presents a snapshot of three timeslots of network performance for a stationary UT experiencing partial obstructions.
Fig.~\ref{subfig:s2_t1} to Fig.~\ref{subfig:s2_t3} illustrate the corresponding cumulative obstruction maps and the connected satellite trajectories at the end of each timeslot.

At \texttt{58}:\texttt{57}, after a regular handover to \texttt{STARLINK-35165}, the UT made a beam switching attempt to \texttt{STARLINK-4709} to mitigate an \texttt{OBSTRUCTED} event resulting from low signal quality.
However, as shown in Fig.~\ref{subfig:s2_t1}, for the remaining duration of this timeslot, the UT continued to experience lower signal quality without making further beam switching attempts.
The downlink throughput performance dropped and continued to degrade at around 50 Mbps until the next regular handover at \texttt{59}:\texttt{12} happened.

In the second timeslot, following the regular handover at \texttt{59}:\texttt{12} to \texttt{STARLINK-5271}, the UT immediately attempted to conduct multiple beam switching to circumvent the \texttt{OBSTRUCTED} event.
Eventually, the UT switched back to \texttt{STARLINK-5271}.
The throughput performance started to recover until the connection to \texttt{STARLINK-5271} was established, with a fluctuated throughput performance below 250 Mbps.
As shown in Fig.~\ref{subfig:s2_t2}, it had a clear LoS without obstructions and strong signal quality for the rest of this timeslot.

In the third timeslot, after the regular handover to \texttt{STARLINK-34280} at \texttt{59}:\texttt{27}, the UT successfully kept to maintain the target bitrate at 250 Mbps without experiencing any outage events so no need to conduct any dynamic beam switching, as shown in Fig.~\ref{subfig:s2_t3} with the clear LoS.

\subsection{Mobile UT}

\begin{figure}
    \centering
    \begin{subfigure}[h]{0.49\linewidth}
        \includegraphics[width=\linewidth]{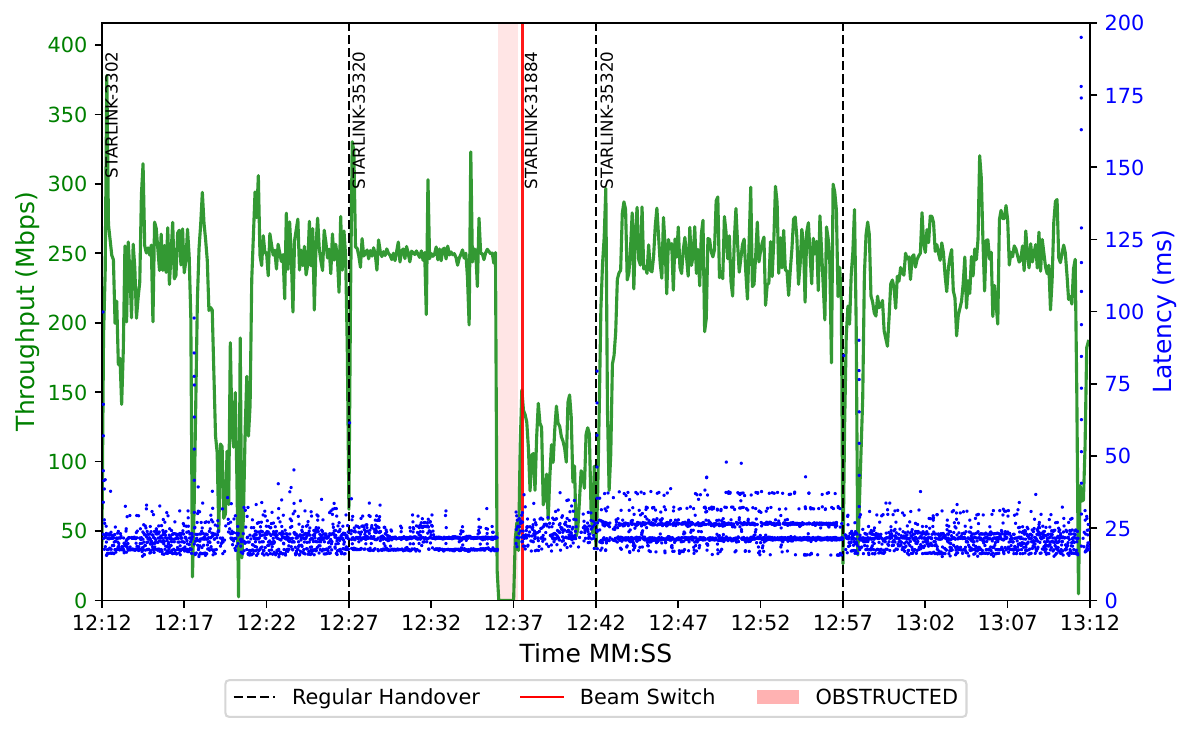}
        \caption{Downlink throughput performance}
        \label{subfig:downlink}
    \end{subfigure}
    \begin{subfigure}[h]{0.49\linewidth}
        \includegraphics[width=\linewidth]{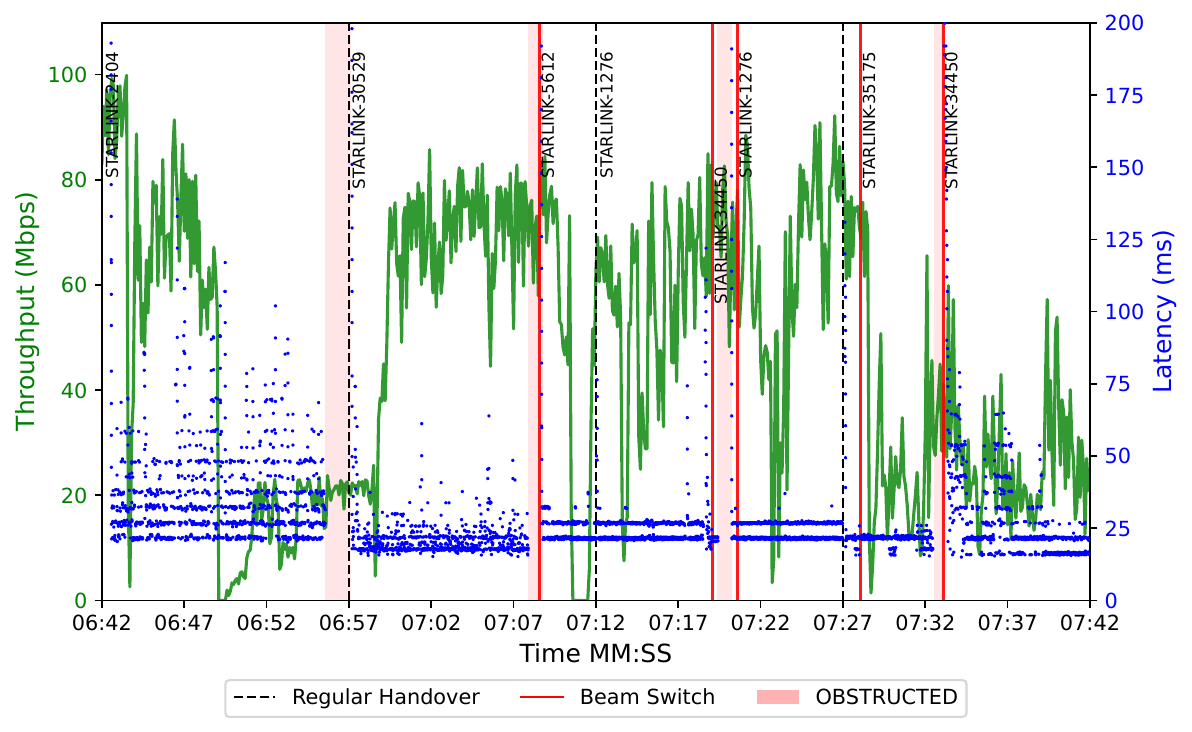}
        \caption{Uplink throughput performance}
        \label{subfig:uplink}
    \end{subfigure}
    \begin{subfigure}[h]{\linewidth}
        \centering
        \includegraphics[width=0.8\linewidth]{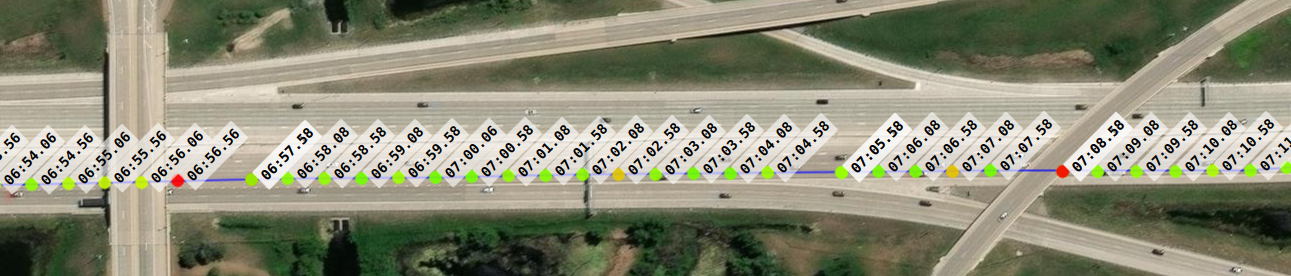}
        \caption{GPS track showing two transient obstructions, causing latency and throughput degradation in (b)}
        \label{subfig:gps-track}
    \end{subfigure}
    \caption{Network performance affected by outage and dynamic beam switching events}
    \label{fig:throughput-outage}
\end{figure}

In Fig.~\ref{fig:throughput-outage}, we present the impact of outage events and dynamic beam switching on network performance for a mobile Starlink UT, with both downlink and uplink throughput performance.

In the first timeslot of Fig.~\ref{subfig:downlink}, we observe a substantial downlink throughput drop (approximately 3~s) caused by a large highway sign.
Although the obstruction map shows a noticeable SNR degradation, the UT maintains its serving beam and does not initiate a beam switch, indicating that the connection remains stable despite the short-term blockage.
In the subsequent timeslot, a regular handover transitions the connection to \texttt{STARLINK-35320}. Shortly after, at \texttt{12}:\texttt{36}, a bridge induces a brief outage (approximately 1~s), which triggers a beam switch to \texttt{STARLINK-31884}.
Following this switch, the downlink throughput drops to roughly half of that observed with \texttt{STARLINK-35320}.
At \texttt{12}:\texttt{42}, when the next regular handover occurs, the UT switched back to \texttt{STARLINK-35320}, and the downlink throughput promptly recovers to its nominal level, on par with the previous timeslot before the outage and beam switch happened.
This suggests that the UT, when forced to conduct a beam switch due to a hard obstruction event, falls back to a non-primary or backup beam in the serving cell with markedly lower performance.

In the first timeslot of Fig.~\ref{subfig:uplink}, we observe a significant uplink throughput drop caused by another large highway sign.
One second before the regular handover at \texttt{06}:\texttt{57}, a bridge-induced LoS obstruction triggers a beam switch, after which the uplink throughput requires (approximately 1~s) to recover during the next timeslot.
A similar pattern appears later in the second timeslot.
At \texttt{07}:\texttt{09}, a bridge caused an outage that triggers a beam switch from \texttt{STARLINK-30529} to \texttt{STARLINK-5612}. As in the downlink case, the uplink throughput does not fully recover until the next regular handover, when the UT returns to a more
favorable serving satellite.
At \texttt{07}:\texttt{12}, a regular handover switches the serving satellite to \texttt{STARLINK-1276}. Immediately after a subsequent outage, the UT briefly switches to \texttt{STARLINK-34550} before returning to \texttt{STARLINK-1276}.
Once the vehicle exits the bridge, the UT re-establishes a stable connection with \texttt{STARLINK-1276}, followed by an approximately 2~s recovery period before throughput returns to nominal levels.
Fig.~\ref{subfig:gps-track} shows the GPS track of both events.

These events reveal a clear asymmetry between schedule-driven regular handovers and dynamic beam switches in LEO satellite networks.
Schedule-driven handovers are pre-coordinated and UT-initiated based on cached satellite schedules, during which packets are briefly bicasted over both the old and new beams~\cite{chenLowLatencyScheduledriven2024a}.
This intentional overlap allows the UT to transition to a high-quality, pre-selected serving beam, often anchored to a planned and well-provisioned ground station, with only short-lived latency spikes and minimal throughput penalty.
In contrast, dynamic beam switches triggered by sudden LoS obstructions are reactive and prioritize link continuity over link quality, without the benefit of packet bicasting or advance coordination.
In these cases, the UT is often reassigned to a sub-optimal backup beam within the current serving cell, which may be anchored to a different or less favorable ground station, potentially introducing additional backhaul latency or congestion.
Because no bicasting window exists, the UT must fully re-establish its link, resulting in prolonged throughput degradation. Notably, even after LoS conditions are restored, the UT does not immediately re-evaluate or return to the optimal beam.
Throughput recovery is frequently delayed until the next scheduled handover, when bicasting and coordinated ground-station selection again become available.

\subsection{FOV Change on Mobile UTs}
\begin{figure*}[tb]
    \centering
    \begin{subfigure}[h]{0.4\linewidth}
        \includegraphics[width=\linewidth]{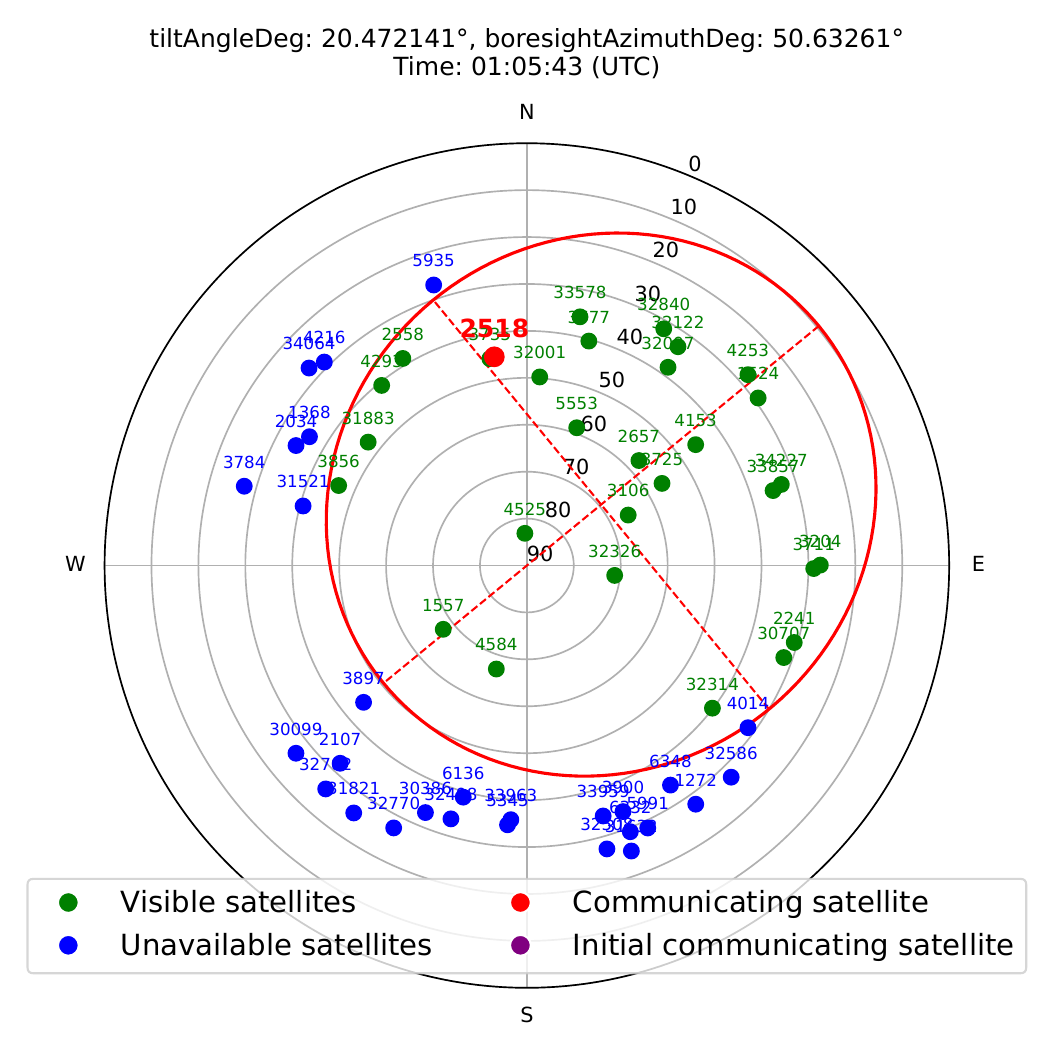}
        \caption{Before rotation}
        \label{subfig:fov_before_rotation}
    \end{subfigure}
    \begin{subfigure}[h]{0.4\linewidth}
        \includegraphics[width=\linewidth]{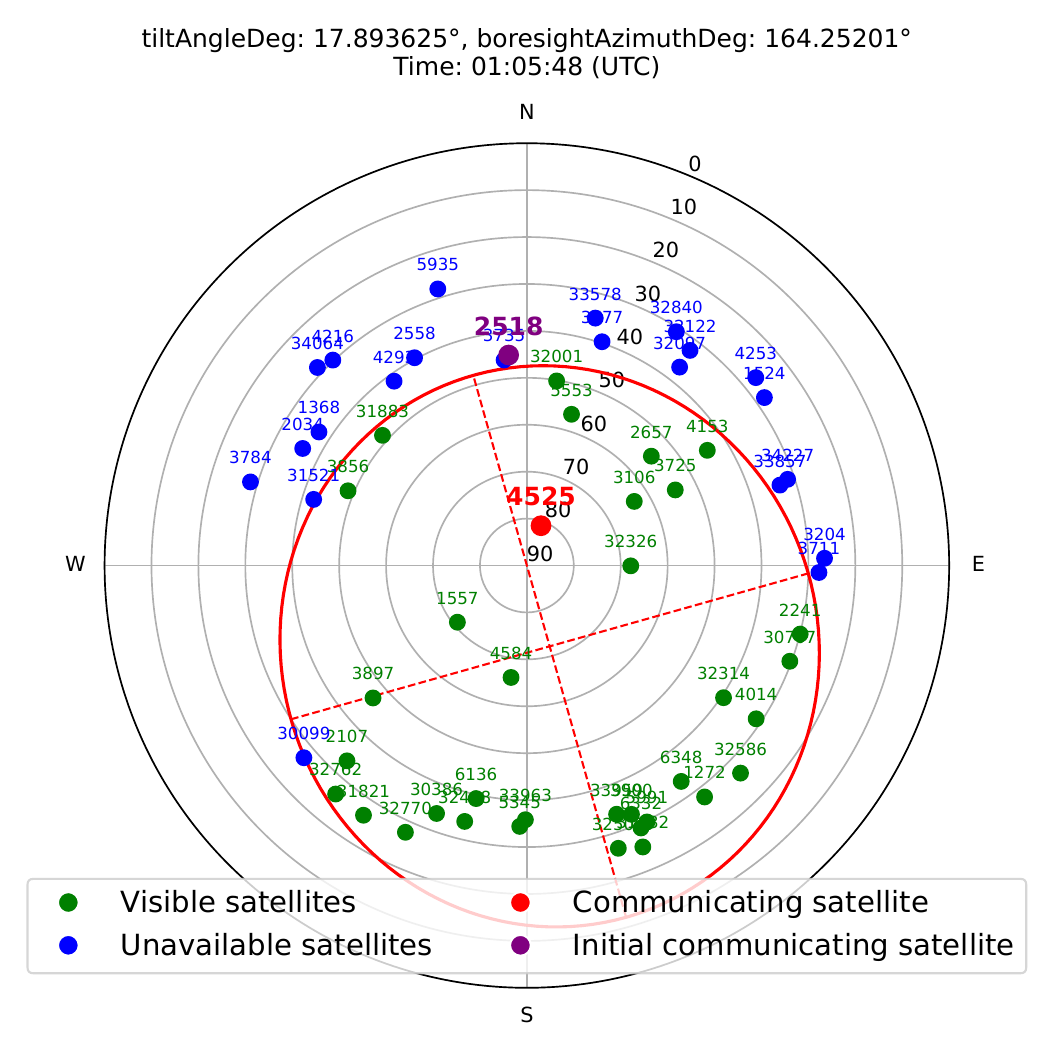}
        \caption{After rotation}
        \label{subfig:fov_after_rotation}
    \end{subfigure}
    \caption{UT's rotational movement changes the FOV and visible satellites. As it rotate clockwise, \texttt{STARLINK-2518} moves out of the FOV, requiring selecting a new satellite within the same timeslot.}
    \label{fig:fov-change-candidate-satellites}
\end{figure*}

Fig.~\ref{fig:fov-change-candidate-satellites} demonstrates how UT mobility affects the FOV and visible satellites.
In this example, the red ellipse represents the UT's FOV, which is calculated based on the specifications released by Starlink\footnote{\url{https://www.starlink.com/specifications}}, with the major and minor axes determined by the UT's current tilt and azimuth.
In Fig.~\ref{subfig:fov_before_rotation}, the UT was connected to \texttt{STARLINK-2518}.
As the UT rotates clockwise, \texttt{STARLINK-2518} moves out of the FOV.
This requires the UT to perform a beam switching event to select a new serving satellite.
\texttt{STARLINK-4525} was selected in Fig.~\ref{subfig:fov_after_rotation}, which had a higher elevation and fit within the FOV.
When the UT is in motion, uneven roads or vehicle turns are more likely to cause the loss of LoS between UTs and communicating satellites, requiring UTs to conduct frequent beam switching events.

\begin{figure}
    \centering
    \begin{subfigure}[h]{0.18\linewidth}
        \includegraphics[width=\linewidth]{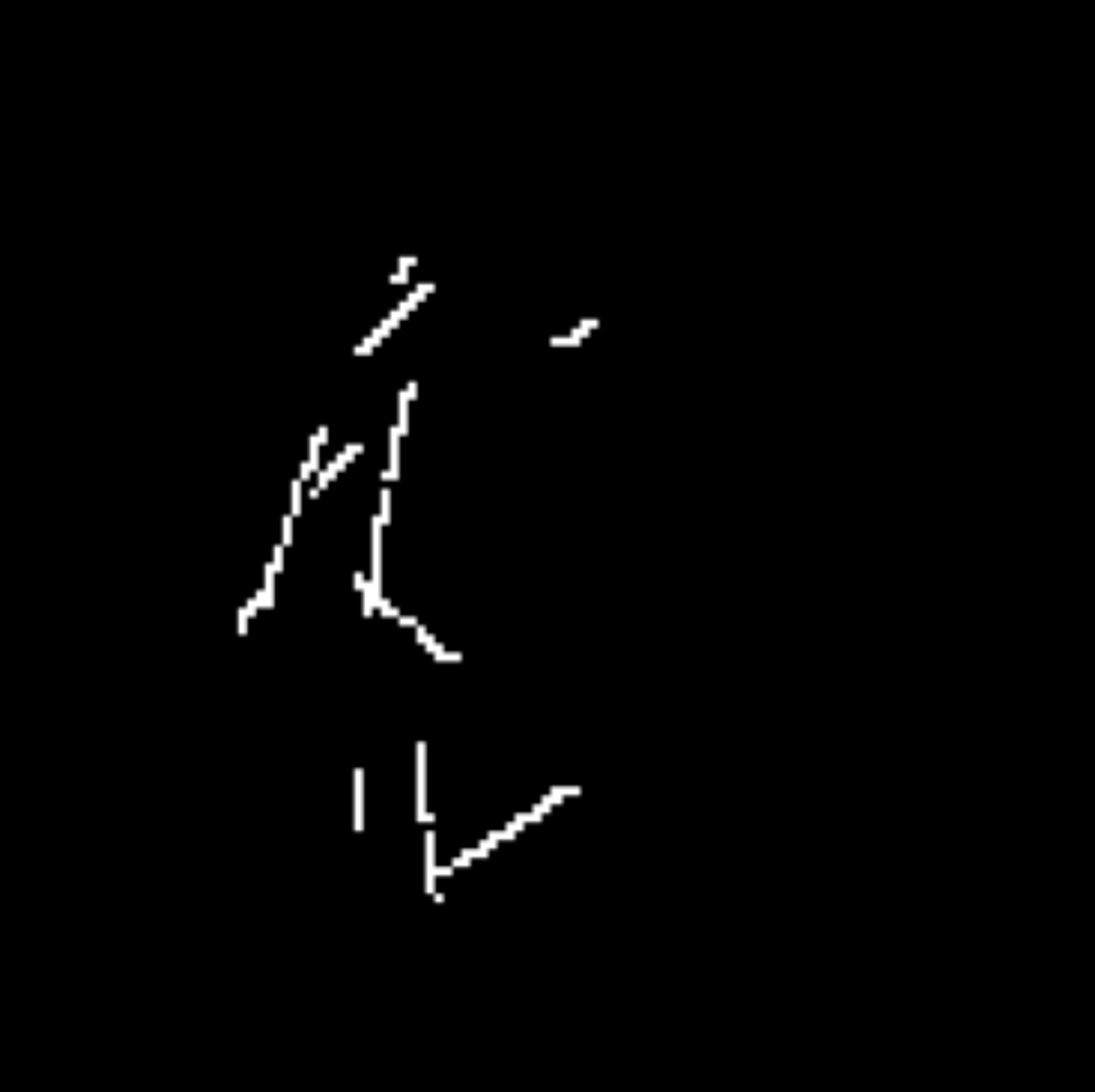}
        \caption{Original obstruction map}
        \label{subfig:cumulative-obstruction-map}
    \end{subfigure}
    \begin{subfigure}[h]{0.18\linewidth}
        \includegraphics[width=\linewidth]{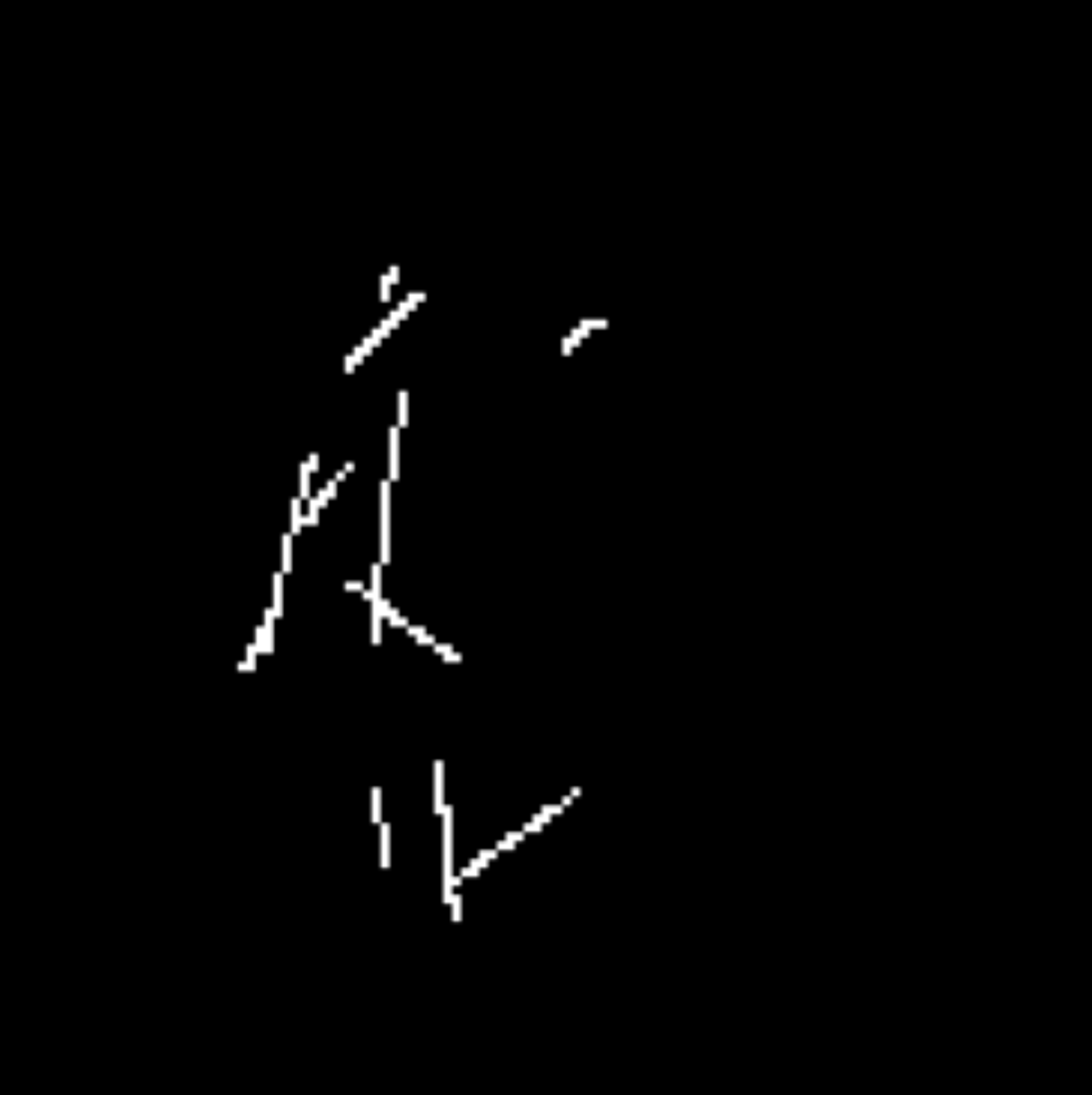}
        \caption{Reconstructed obstruction map}
        \label{subfig:reconstructed-obstruction-map}
    \end{subfigure}
    \caption{Comparison between observed and reconstructed obstruction maps}
    \label{fig:obstruction-map-comparison}
\end{figure}

\subsection{Satellite Identification Accuracy}

To verify the accuracy of satellite identification results, particularly when the UT is in motion, we use the SGP4~\cite{sgp4} algorithm to calculate the expected satellite ground track for a specific Starlink satellite ID at a given time.
We then reconstruct the anticipated obstruction map by incorporating the UT's location, alignment, and orientation to determine the satellite's azimuth and elevation relative to the UT, projecting this direction onto the obstruction map.
We compare the reconstructed obstruction maps with the observed ones retrieved from the gRPC interface to validate the identification results, as shown in Fig.~\ref{fig:obstruction-map-comparison}.
The satellite trajectories of connected satellites in the reconstructed obstruction map, as shown in Fig.~\ref{subfig:reconstructed-obstruction-map}, closely matches the observed obstruction map, with 1--2 pixel differences primarily due to rounding errors when projecting the satellite's azimuth and elevation onto the obstruction map with a resolution of 123x123 pixels.

To evaluate the accuracy of the proposed mobility-aware satellite identification algorithm with quantitative metrics, we further calculate the angular separation between candidate satellite locations projected from the 2D obstruction map and the identified satellites using our method.
In stationary scenarios, the average angular separation is 2.37$^\circ$, with a standard deviation of 1.73$^\circ$.
In our mobility measurement, the average angular separation slightly increases to 5.44$^\circ$, with a standard deviation of 3.64$^\circ$.

\section{Limitations}\label{sec:limitations}

With over 9,000 operational satellites in orbit as of December 2025 and the continuous deployment of new satellites, the density of Starlink satellites across various orbital shells is steadily increasing.
Consequently, with the limited spatial resolution of 123x123 pixels in obstruction maps, it is becoming more challenging to distinguish adjacent satellites with similar azimuth and elevation based on their projected pixel locations.
This could affect the accuracy of identifying the exact communicating satellite IDs for UTs.
However, considering the satellite trajectories and their relative elevations to UTs, it may be feasible to cluster regions of adjacent satellites with similar performance characteristics, as they often follow similar paths and might connect to nearby ground stations.
Unlike other LEO network operators such as OneWeb, without Starlink exposing the exact connected satellite IDs, obstruction map-based methods remain the only feasible option for identifying communicating satellites for Starlink UTs.
We acknowledge Starlink's efforts to increase transparency by updating comprehensive event logs\footnote{\url{https://starlink.com/support/article/0d8ed4aa-e19c-919f-4e0c-93c12656fc48}}.
We hope that future firmware updates will restore useful diagnostic information, such as connected satellite IDs, connected ground stations, or physical layer metrics such as SNR or RSSI, to facilitate the development of LEO-aware transport protocols and applications, further improving the overall user experience.

\section{Discussion and Future Work}\label{sec:discussion}

\noindent
\textbf{Network Performance Improvements}: In some mobility scenarios, vehicles have fixed routes, such as fixed route public transit or Starlink UTs installed on railways.
Normally, due to the unpredictable vehicle movements, it is not feasible to predict transient obstructions and the impact on network performance.
With a recurring fixed route, it is possible to utilize historical network measurement data to learn the performance pattern, and optimize for upper layer protocols and applications, such as multipath packet scheduling on vehicles equipped with both Starlink and cellular services, to conceal the impact of transient obstructions for transport layer protocols and various applications.

\noindent
\textbf{Generalization to Other LEO Satellite Networks}:
While our study focuses on Starlink UTs, other LEO satellite networks, such as OneWeb, also support mobility and maritime use cases~\cite{onewebroadrailway,onewebmaritime}.
Given that OneWeb employs a ``Walker-Star'' constellation topology with around 650 satellites across 12 near-polar orbital planes, along with different beam allocation and handover strategies than Starlink~\cite{zhaoMeasuringOneWebSatellite2025}, it would be valuable to investigate how OneWeb UTs perform in mobility scenarios with transient obstructions and rapidly changing FOVs.

\section{Related Works}\label{sec:related}

\noindent
\textbf{Starlink Satellite Identification.}
Researchers have applied various approaches to measure the performance of Starlink networks in stationary settings, such as evaluating the latency performance~\cite{izhikevichDemocratizingLEOSatellite2024b,zhaoLENSLEOSatellite2024b,panMeasuringSatelliteLinks2024b}, modeling and predicting the downlink network throughput~\cite{ukwenExaminingPredictabilityStarlink2024,garciaInferringStarlinkPhysical2024b,garciaModelingPredictingStarlink2025}.
To understand the disparities in network measurement results, researchers have started utilizing the diagnostic information offered by the gRPC interface for additional insights.
Hammas et al.~\cite{tanveerMakingSenseConstellations2023c} first investigated the potential capability of using obstruction maps and satellite ephemeris with the Two-Line element (TLE) data to identify communicating satellites.
Their approach required manually rebooting the UT to trigger the reset and reconstruction of obstruction maps at the time of their work.
They did not provide a practical implementation, as their satellite identification method was based on certain simplified assumptions.
Ahangarpour et al.~\cite{ahangarpourTrajectorybasedServingSatellite2024b} conducted the first systematic study of obstruction maps, leveraging the new capability of the gRPC interface that allows manually triggering the reconstruction of the obstruction map without rebooting the UT.
They presented a practical implementation capable of accurately identifying communicating satellites in stationary scenarios.
Liu et al.~\cite{liuVivisectingStarlinkThroughput2025} proposed similar techniques to identify communicating satellites in stationary scenarios and developed a learning-based model to predict downlink throughput using identified satellite IDs, historical data, and environmental factors such as weather.
However, their satellite identification approach did not account for varying reference frames in obstruction maps or FOV changes due to UT orientation, even in stationary conditions, as it assumed a fixed radius of 52 pixels for all obstruction maps.
Additionally, they did not explicitly study the impact of dynamic beam switching, or utilize the event logs from the gRPC interface to provide a physically plausible explanation for performance degradations.

\noindent
\textbf{Mobile Starlink Performance.}
In recent years, several studies~\cite{huLEOSatelliteVs2023a,laniewskiStarlinkRoadFirst2024,beckmanStarlinkCellularConnectivity2024,laniewskiMeasuringMobileStarlink2025,wangExploring5GDigital2025} have started to explore the Starlink network performance in mobility settings.
Hu et al.~\cite{huLEOSatelliteVs2023a} compared the mobile Starlink performance with major cellular providers in the USA, focusing solely on general network measurement metrics such as latency, throughput, and packet loss.
Beckman et al.~\cite{beckmanStarlinkCellularConnectivity2024} conducted similar measurements and comparisons in Northern Europe, investigating the impact of different satellite orbital inclinations and the reduced satellite density while traversing the Arctic Circle.
Laniewski et al.~\cite{laniewskiStarlinkRoadFirst2024,laniewskiMeasuringMobileStarlink2025} provided initial discussions on how transient obstructions affect mobile Starlink performance.
Their test drive was conducted on a highway, offering an almost optimal, obstruction-free FOV for the UT throughout the measurements.
Consequently, they primarily concentrated on scenarios involving bridges crossing highways to study the impact of transient obstructions on network performance.
Their analysis confirmed that such transient obstructions can cause increased packet loss, latency spikes, and reduced throughput.
Overall, they observed approximately 3\% of total packet loss during their 300 km highway test drives.
Among all packet loss events, they attributed 32\% to the regular 15-second handover behavior and transient obstructions caused by bridges.
However, they did not provide detailed insights into the remaining packet loss events, which could potentially be attributed to the loss of LoS or changes in FOV.
Wang et al.~\cite{wangExploring5GDigital2025} compared the network performance of Starlink and 5G networks in the non-contiguous U.S., specifically in Alaska and Hawaii.
Their measurements focused primarily on network performance metrics such as latency and throughput.
While they provided high-level statistics on the percentage of outage time reported from the Starlink gRPC interface, in urban and rural areas during the overall test drive, they did not delve into the specific causes of these outages.

\noindent
\textbf{Starlink Handover Dynamics.}
Some recent works~\cite{laiMindMisleadingEffects2024,laiLeoCCMakingInternet2025a} have conceptually introduced aperiodic satellite handovers, occurring in addition to regular 15-second global handovers, attributing them to factors such as poor SNR or obstructions, which lead to variations in network performance in terms of latency and throughput.
Liu et al.~\cite{liuVivisectingStarlinkThroughput2025} further discussed not all regular handovers would result in the change of connected satellites, as long as the previously connected satellite remained in the FOV in obstruction-free stationary scenarios.
However, latency spikes and throughput drops can still occur during these regular handovers, indicating a potential inter-beam handover involving the same satellite.

\section{Ethical Considerations}

This work does not have any ethical concerns.

\section{Conclusion}\label{sec:conclusion}

In this paper, we introduce a mobility-aware algorithm to identify communicating satellites for Starlink UTs in motion.
Beyond the regular schedule-driven 15-second handover intervals, we revealed that Starlink UTs can perform multiple beam switching attempts to connect to different satellites, circumventing transient obstructions or degraded LoS conditions.
We demonstrated the network impact of beam switching events, especially prolonged connection outages and the occurrence of latency spikes in addition to regular handover events.
By identifying the communicating satellites for Starlink UTs in motion, this paper paves the way for further research into understanding the end-to-end performance of Starlink networks and its satellite selection strategies.
Future research on developing LEO constellation simulations can incorporate both the communicating satellite identification results and network performance traces for realistic and mobility-aware scenarios.

\pagebreak

\bibliographystyle{ACM-Reference-Format}
\bibliography{reference}

\appendix

\section{Starlink Outage Event Statistics}\label{sec:outage-event-statistics}

\begin{table}[h]
\caption{Outage event statistics in mobile and stationary obstructed scenarios}
\label{table:outage-event-statistics}
\begin{tabular}{|l|ll|ll|}
\hline
\multirow{2}{*}{Cause} & \multicolumn{2}{l|}{Time (s)}             & \multicolumn{2}{l|}{\% of Outages}       \\ \cline{2-5}
                       & \multicolumn{1}{l|}{Mobile}  & Stationary & \multicolumn{1}{l|}{Mobile} & Stationary \\ \hline
NO\_DOWNLINK           & \multicolumn{1}{l|}{133.958} & 132.265    & \multicolumn{1}{l|}{27.333}  & 20.216     \\ \hline
NO\_PINGS              & \multicolumn{1}{l|}{23.659}  & 5.999      & \multicolumn{1}{l|}{4.827}   & 0.917       \\ \hline
OBSTRUCTED             & \multicolumn{1}{l|}{110.858} & 367.764    & \multicolumn{1}{l|}{22.620}  & 56.212      \\ \hline
SKY\_SEARCH            & \multicolumn{1}{l|}{221.404} & 148.222    & \multicolumn{1}{l|}{45.175}  & 22.655      \\ \hline
UNKNOWN                & \multicolumn{1}{l|}{0.220}   & 0          & \multicolumn{1}{l|}{0.0449}    & 0        \\ \hline
\end{tabular}
\end{table}

\section{Starlink UT Event Reasons}

\begin{table}[h]
\caption{Starlink UT event reasons}
\label{tab:ut-outage-reasons}
\begin{tabular}{l}
\hline
\textbf{Event Reason}   \\ \hline
EVENT\_REASON\_OUTAGE\_UNKNOWN            \\ \hline
EVENT\_REASON\_OUTAGE\_BOOTING            \\ \hline
EVENT\_REASON\_OUTAGE\_STOWED             \\ \hline
EVENT\_REASON\_OUTAGE\_THERMAL\_SHUTDOWN  \\ \hline
EVENT\_REASON\_OUTAGE\_NO\_SCHEDULE       \\ \hline
EVENT\_REASON\_OUTAGE\_NO\_SATS           \\ \hline
EVENT\_REASON\_OUTAGE\_OBSTRUCTED         \\ \hline
EVENT\_REASON\_OUTAGE\_NO\_DOWNLINK       \\ \hline
EVENT\_REASON\_OUTAGE\_NO\_PINGS          \\ \hline
EVENT\_REASON\_OUTAGE\_SLEEPING           \\ \hline
EVENT\_REASON\_OUTAGE\_MOVING\_WHILE\_NOT\_ALLOWED     \\ \hline
EVENT\_REASON\_OUTAGE\_SKY\_SEARCH                     \\ \hline
EVENT\_REASON\_HIGH\_DOWNLINK\_PACKET\_LOSS            \\ \hline
EVENT\_REASON\_UT\_ALERT\_RAIN\_SNR\_PERSISTENTLY\_LOW \\ \hline
EVENT\_REASON\_UT\_ALERT\_ETH\_NO\_LINK                \\ \hline
EVENT\_REASON\_UT\_ALERT\_ETH\_SLOW\_LINK              \\ \hline
EVENT\_REASON\_UT\_ALERT\_ETH\_SLOW\_LINK\_100         \\ \hline
\end{tabular}
\end{table}

As of Starlink UT firmware version \texttt{2025.12.02.mr68972.1}, there are 17 distinct event reasons related to UT outages, as shown in Table~\ref{tab:ut-outage-reasons}.
This list is decoded from Starlink UT's gRPC protobuf definitions.

\section{Starlink UT gRPC Data Fields}

\begin{table}[h]
\caption{Fields collected for the UT status}
\centering
\begin{tabular}{l l}
\hline
\textbf{Attribute} & \textbf{Type} \\
\hline
timestamp & float (Unix epoch) \\
hardware\_version & string \\
pop\_ping\_latency\_ms & float \\
downlink\_throughput\_bps & float \\
uplink\_throughput\_bps & float \\
tilt\_angle\_deg & float \\
boresight\_azimuth\_deg & float \\
boresight\_elevation\_deg & float \\
attitude\_estimation\_state & int (enum) \\
attitude\_uncertainty\_deg & float \\
desired\_boresight\_azimuth\_deg & float \\
desired\_boresight\_elevation\_deg & float \\
q\_scalar, q\_x, q\_y, q\_z & float \\
\hline
\end{tabular}
\end{table}

\begin{table}[h]
\caption{Fields collected for the UT location}
\centering
\begin{tabular}{l l}
\hline
\textbf{Attribute} & \textbf{Type} \\
\hline
timestamp & float (GPS epoch) \\
latitude & float \\
longitude & float \\
altitude\_meters & float \\
horizontal\_speed\_mps & float \\
vertical\_speed\_mps & float \\
source & string \\
\hline
\end{tabular}
\end{table}

\begin{table}[h]
\caption{Fields collected for the UT obstruction map}
\centering
\begin{tabular}{l l}
\hline
\textbf{Attribute} & \textbf{Type} \\
\hline
timestamp & float (Unix epoch) \\
frame\_type & int (enum: \texttt{FRAME\_EARTH} or \texttt{FRAME\_UT}) \\

raw & float matrix (123$\times$123) \\
\hline
\end{tabular}
\end{table}

\end{document}